\documentclass[aps,twocolumn,prx,preprintnumbers,showpacs,amsmath,amssymb,superscriptaddress]{revtex4-1}
\usepackage[pdftex, colorlinks, citecolor=blue]{hyperref}   
\usepackage{graphicx}
\usepackage{dcolumn}
\usepackage{threeparttable}
\usepackage{multirow}
\usepackage{booktabs}
\usepackage{txfonts}
\usepackage{xcolor}
\usepackage{bm}
\usepackage{amssymb}
\usepackage{amsmath}
\usepackage{latexsym}
\usepackage{epsfig}
\usepackage{amsbsy}
\usepackage{array}
\usepackage{tabularx}
\usepackage{esvect} 
\usepackage{extarrows}

\newcommand{\bra}[1]{\langle #1 \mid}
\newcommand{\ket}[1]{\mid #1 \rangle}

\newcommand{\RR}{\mathbf{R}}
\newcommand{\rr}{\mathbf{r}}
\newcommand{\kk}{\mathbf{k}}
\newcommand{\ii}{\mathrm{i}}

\newcommand{\nk}{{n\mathbf{\kk}}}

\begin{document}

\title{Assessment of electric-field-induced forces for bulk solids \\
within the projector-augmented wave method}

\author{Peitao Liu}
\email{ptliu@imr.ac.cn}
\affiliation{Shenyang National Laboratory for Materials Science, Institute of Metal Research, Chinese Academy of Sciences, 110016 Shenyang, China}

\author{Martijn Marsman}
\affiliation{University of Vienna, Faculty of Physics and Center for Computational Materials Science, Kolingasse 14-16, A-1090 Vienna, Austria}

\author{Georg Kresse}
\affiliation{University of Vienna, Faculty of Physics and Center for Computational Materials Science, Kolingasse 14-16, A-1090 Vienna, Austria}

\author{Cesare Franchini}
\affiliation{University of Vienna, Faculty of Physics and Center for Computational Materials Science, Kolingasse 14-16, A-1090 Vienna, Austria}
\affiliation{Dipartimento di Fisica e Astronomia, Universit\`{a} di Bologna, 40127 Bologna, Italy}

\begin{abstract}
We assess the performance of electric-field-induced forces for bulk solids within the projector-augmented wave method.
This capability allows for first-principles structural relaxations of bulk semiconductors and insulators under external electric fields,
thereby enabling the calculation of linear and nonlinear responses to electric fields as well as properties associated with electric-field-driven structural changes.
The approach is first validated by presenting calculated ion-clamped and unclamped dielectric and piezoelectric constants, quadratic dielectric susceptibilities,
and Born effective charges for semiconductors AlX (X=P, As and Sb), showing good agreement with density functional perturbation theory calculations
as well as with available experimental and theoretical data.
As a further demonstration, we examine the electric-field-driven antiferroelectric-to-ferroelectric phase transition in AgNbO$_3$,
for which our computed polarization versus electric field hysteresis loop aligns with previous calculations.
\end{abstract}

\maketitle

\section{Introduction}\label{sec:intro}

Electric fields are commonly used in experiments for manipulating a variety of properties, e.g.,
magnetic interactions~\cite{Ohno_Nature2000,Zhao_NM_2006, Chiba_Nature2008,Heron_Nature_2014,Gao_NanoLett2017,PhysRevLett.107.217202}
(see reviews in Refs.~\cite{doi:10.1063/1.4870957, SONG201733_review,Spaldin_NR_2019}),
exchange bias~\cite{Wu_NM_2010,He_NM_2010}, metal-insulator transition~\cite{doi:10.1002/adma.201003241},
superconductivity~\cite{Ueno2008,Caviglia_2008}, optical absorption~\cite{NEWAZ201349},
and structural distortions~\cite{doi:10.1063/1.4898573,PhysRevLett.120.017201,Yi_NC2020}.
However, the nonperiodic nature of the electric-field potential makes its treatment nontrivial within first-principles methods based on periodic boundary conditions~\cite{RevModPhys.63.91}.
In addition, the unboundedness of the electric field potential can always lower the energy
of an insulating system by transferring charge from valence bands of one region to
conduction bands in another distant region, resulting in a dielectric breakdown~\cite{RevModPhys.63.91,KANE1960181}.
Nevertheless, under a small finite-electric-field when the Zener tunneling is not severe,
it turned out that the density matrix remains periodic under translation by a lattice vector~\cite{PhysRevLett.89.117602}
and the long-lived field-polarized states can be obtained by minimizing the electric enthalpy
functional~\cite{PhysRevB.63.155107,PhysRevLett.89.117602,PhysRevLett.89.157602,PhysRevB.66.104108}
with the electronic polarization calculated using the modern theory of polarization~\cite{PhysRevB.47.1651,RevModPhys.66.899}.

The first-principles finite-electric-field approach was first implemented independently
by Souza, \'{I}\~{n}guez, and Vanderbilt~\cite{PhysRevLett.89.117602}
and Umari and Pasquarello~\cite{PhysRevLett.89.157602} using norm-conserving pseudopotentials,
and was later extended to ultrasoft pseudopotentials by Vanderbilt and King-Smith~\cite{David1998_USPP} and
the projector augmented wave (PAW) pseudopotentials by Zwanziger \emph{et al.}~\cite{ZWANZIGER2012113}.
The approach not only allows for straightforward calculation of various responses
of insulators and semiconductors to external electric fields, e.g.,
linear and nonlinear dielectric susceptibilities~\cite{PhysRevB.71.125107,PhysRevB.74.245204},
piezoelectric constants~\cite{PhysRevLett.62.2853,MinamiNakayamaShimada2024}, Born effective charges~\cite{PhysRevLett.89.117602,PhysRevLett.89.157602,PhysRevB.75.115116},
longitudinal nonlinear conductivity~\cite{PhysRevLett.134.046801},
Hyper-Raman spectra~\cite{PhysRevLett.98.176402}, and Zeeman spin splitting~\cite{96gy-sn83}, but also enables the study of
the associated electric-field-driven structural changes~\cite{PhysRevB.66.104108,doi:10.1063/1.4941319,PhysRevLett.91.057601,bilc2016highfieldpropertiestypical,7939087,
GHOSH2019166,Chen2025,PhysRevLett.116.057602,PhysRevLett.122.247701,ZhangEtAl2022KNN,ArrasPaillardBellaiche2023,HanEtAl2024AFEFE}.

With the growing use of electric fields as a versatile tool for tuning
material properties in both experiments and simulations, robust and efficient
approaches for treating finite-electric-fields in bulk solids are increasingly
valuable for the \emph{ab initio} community.
A recent review of first-principles modeling of ferroelectric oxide perovskites
discusses the modern theory of polarization, explicit finite-electric-field
methods, and the calculation of functional and electromechanical responses~\cite{GhosezJunquera2022}.
While several well-known packages, such as ABINIT~\cite{GONZE2002478,PhysRevLett.89.117602,ZWANZIGER2012113},
Quantum ESPRESSO~\cite{QE-2009,PhysRevLett.89.157602}, and CASTEP~\cite{CASTEP},
already incorporate finite-electric-field capabilities, the present work
provides a systematic assessment of the electric-field-induced forces for bulk solids within
the PAW formalism as implemented in the Vienna \emph{Ab initio} Simulation
Package (VASP)~\cite{PhysRevB.47.558, PhysRevB.54.11169}.
Specifically, we benchmark the accuracy and reliability of the finite-electric-field
response by evaluating ion-clamped and relaxed-ion dielectric and
piezoelectric constants, quadratic dielectric susceptibility, and Born
effective charges for the prototypical semiconductors AlX (X=P, As, and Sb).
The obtained results show good agreement with density-functional perturbation
theory (DFPT) calculations and available experimental and theoretical data.
As a further validation, we reproduce the electric-field-driven
antiferroelectric-to-ferroelectric phase transition in AgNbO$_3$~\cite{doi:10.1063/1.4941319}.
The paper is organized as follows. In Section~\ref{sec:method}, we detail the
methodology underlying the finite-electric-field treatment as implemented in
VASP. Computational details are provided in Section~\ref{sec:tech}.
The benchmark results are presented and discussed in Section~\ref{sec:results},
and a summary is given in Section~\ref{sec:conclusions}.

\section{Method}\label{sec:method}

The theoretical ingredients underlying finite-electric-field calculations have been established in several previous works, including the electric-enthalpy formulation and Berry-phase treatment of polarization~\cite{PhysRevB.63.155107,PhysRevLett.89.117602,PhysRevLett.89.157602,PhysRevB.66.104108}, as well as their extensions to ultrasoft pseudopotentials and the PAW framework~\cite{David1998_USPP,ZWANZIGER2012113}. Rather than introducing a new theoretical framework, the purpose of this section is to provide a comprehensive, structured, and self-contained summary of the formalism relevant to the electric-field-induced forces assessed in the present work. This presentation also establishes the notation and theoretical basis needed for the subsequent discussion and benchmarking of the VASP implementation.

For the derivations in this section, we follow the conventional Hartree atomic-unit notation used in the finite-electric-field literature,
while retaining the elementary charge $e$ explicitly for clarity.
For the evaluation and reporting of dielectric, nonlinear dielectric, and piezoelectric response coefficients in Section~\ref{sec:tech},
the polarization and electric field are converted to SI units.
Accordingly, the SI constitutive relation $\mathbf{D}=\epsilon_0\boldsymbol{\mathcal E}+\mathbf{P}$ is used there.

We first summarize the finite-electric-field formalism in the all-electron representation in Section~\ref{sec:EFIELD_all-electron}, and then describe its extension to the PAW framework in Section~\ref{sec:EFIELD_PAW}.

\subsection{Electric field in the all-electron formalism}\label{sec:EFIELD_all-electron}

The long-lived electronic state of an insulating system under a finite homogeneous electric
field $\boldsymbol{\mathcal E}$ is determined by solving for the field-polarized
Bloch functions $\{\psi^{(\boldsymbol{\mathcal E})}\}$ through minimization of the electric
enthalpy functional~\cite{PhysRevB.63.155107,PhysRevLett.89.117602,PhysRevLett.89.157602,PhysRevB.66.104108}
\begin{equation}\label{eq:Efield_functional}
\mathcal{F}[\{\psi^{(\boldsymbol{\mathcal E})}\},\boldsymbol{\mathcal E}]
=
E^{(\boldsymbol{\mathcal E})}_{\rm KS}
[\{\psi^{(\boldsymbol{\mathcal E})}\}]
-\Omega \boldsymbol{\mathcal E}\cdot
{\bf P}_{\rm mac}
[\{\psi^{(\boldsymbol{\mathcal E})}\}].
\end{equation}
Here, $\Omega$ is the unit-cell volume.
$E^{(\boldsymbol{\mathcal E})}_{\rm KS}$ denotes the Kohn-Sham (KS) total-energy
functional evaluated for the field-polarized electronic state. It can be written in terms
of the KS single-particle contribution as
\begin{equation}\label{eq:KS_energy}
E^{(\boldsymbol{\mathcal E})}_{\rm KS}
[\{\psi^{(\boldsymbol{\mathcal E})}\}]
=
\frac{f}{N_{\kk}}
\sum_{\kk}\sum_n^{\rm occ}
\left\langle
u^{(\boldsymbol{\mathcal E})}_{n\kk}
\left|
H^{(\kk)}_{\rm KS}
\right|
u^{(\boldsymbol{\mathcal E})}_{n\kk}
\right\rangle
+
E_{\rm dc}^{(\boldsymbol{\mathcal E})},
\end{equation}
where $u^{(\boldsymbol{\mathcal E})}_{n\kk}$ is the cell-periodic part of
$\psi^{(\boldsymbol{\mathcal E})}_{n\kk}$,
$H^{(\kk)}_{\rm KS}$ is the KS Hamiltonian, $f$ is the spin degeneracy,
$\kk$ and $n$ are the $k$-point and band indices, respectively, and
$N_{\kk}$ is the number of $k$ points used to sample the Brillouin zone (BZ).
$E_{\rm dc}^{(\boldsymbol{\mathcal E})}$ denotes the usual double-counting corrections required to recover the KS total energy
from the sum of single-particle expectation values. Although $E_{\rm dc}^{(\boldsymbol{\mathcal E})}$ depends implicitly on the
occupied orbitals through the electron density, its variation, together with the density dependence of the KS Hamiltonian in the
single-particle term, yields the usual KS functional derivative,
\begin{equation}
\frac{\delta E_{\rm KS}^{(\boldsymbol{\mathcal E})}}
{\delta\langle u_{n\mathbf{k}}^{(\boldsymbol{\mathcal E})}|}
=
\frac{f}{N_{\mathbf{k}}}
H_{\rm KS}^{(\mathbf{k})}
|u_{n\mathbf{k}}^{(\boldsymbol{\mathcal E})}\rangle .
\end{equation}

The macroscopic polarization ${\bf P}_{\rm mac}$ contains both ionic
(${\bf P}_{\rm ion}$) and electronic (${\bf P}_{\rm el}$) contributions,
\begin{equation}
{\bf P}_{\rm mac}
[\{\psi^{(\boldsymbol{\mathcal E})}\}]
=
{\bf P}_{\rm ion}
+
{\bf P}_{\rm el}
[\{\psi^{(\boldsymbol{\mathcal E})}\}].
\end{equation}
The ionic contribution is given by
\begin{equation}
\mathbf{P}_{\rm ion}
=
\frac{e}{\Omega}
\sum_{\tau=1}^{N_{\rm ion}}
Z^{\rm ion}_\tau
\mathbf{R}_\tau,
\end{equation}
where $e$ is the positive elementary charge,
$N_{\rm ion}$ is the number of ions in the unit cell,
$Z^{\rm ion}_\tau$ is the ionic charge of ion $\tau$ in unit of $e$,
and $\mathbf{R}_\tau$ is the position vector of ion $\tau$ with respect to
the chosen origin of the unit cell. The choice of origin only affects the polarization branch, but does not alter the polarization differences relevant for finite-field response calculations.
The electronic contribution
${\bf P}_{\rm el}[\{\psi^{(\boldsymbol{\mathcal E})}\}]$
is calculated by integrating the Berry connection around a circuit in the BZ
relying on the modern theory of polarization~\cite{PhysRevB.47.1651,RevModPhys.66.899}
\begin{equation}
{\bf P}_{\rm el}[\{\psi^{(\boldsymbol{\mathcal  E})}\}]=-\frac{\ii fe}{(2\pi)^3}\sum_n^{\rm occ}
\int_{\rm BZ}  \langle u^{(\boldsymbol{\mathcal  E})}_{n{\bf k}}|{\bf
\nabla_k}| u^{(\boldsymbol{\mathcal  E})}_{n{\bf k}} \rangle d{\bf k}.
\end{equation}

In practice, the electronic polarization is calculated by summation over a discrete $k$-point mesh~\cite{PhysRevLett.89.117602, PhysRevB.47.1651}
\begin{equation}\label{eq:P_BP_1}
\begin{split}
{\bf P}_{\rm el} \cdot \mathbf{b}_i
= \frac{fe}{\Omega} \varphi^{(i)}_{\rm el},
\end{split}
\end{equation}
where $\varphi^{(i)}_{\rm el}$ is the string-averaged discretized Berry phase along the direction of the reciprocal lattice vector $\mathbf{b}_i $
\begin{equation}\label{eq:P_BP_2}
\begin{split}
\varphi^{(i)}_{\rm el}
= \frac{1}{N^{(i)}_{k_{\perp}}}\sum_{\kk^{(i)}_{\perp}}^{N^{(i)}_{k_{\perp}}}
 \text{Im} \Bigg\{\ln \prod^{N^{(i)}_{k_{\parallel}}-1}_{j=0} \text{det} \left( M^{(\kk_j, \kk_{j+1})}_{mn} \right)\Bigg\},
\end{split}
\end{equation}
where $N^{(i)}_{k_{\perp}}$ is the number of $k$-points (or strings) in the plane perpendicular to $\mathbf{b}_i $
and $\kk_j= \kk^{(i)}_{\perp} + j (\mathbf{b}_i/N^{(i)}_{k_{\parallel}}) $ is the $j$-th $\kk$ point
in the string of $\kk^{(i)}_{\perp}$ with $N^{(i)}_{k_{\parallel}}$ being the number of $k$ points in the string.
$M^{(\kk_j, \kk_{j+1})}_{mn}$ is the overlap matrix defined by
\begin{equation}\label{eq:overlap_all_electron}
M^{(\kk_j, \kk_{j+1})}_{mn}= \langle u^{(\boldsymbol{\mathcal  E})}_{m{\bf k}_{j}}|u^{(\boldsymbol{\mathcal  E})}_{n{\bf k}_{j+1}}\rangle,
\end{equation}
where $m$ and $n$ enumerate the $N$ occupied bands.

With the electric enthalpy functional in Eq.~\eqref{eq:Efield_functional}, the gradient is then obtained by
\begin{equation}\label{eq:gradient}
\frac{\delta \mathcal{F}[\{u^{(\boldsymbol{\mathcal  E})}\},\boldsymbol{\mathcal  E}]}{\delta \langle u^{(\boldsymbol{\mathcal  E})}_{n\mathbf{k}}|}=
\frac{f}{N_\kk} H^{(\kk)}_{\rm KS} |u^{(\boldsymbol{\mathcal  E})}_{n{\bf k}}\rangle
-\Omega \boldsymbol{\mathcal  E}\cdot
\frac{\delta {\bf P}_{\rm el}[\{u^{(\boldsymbol{\mathcal  E})}\}]}
{\delta \langle u^{(\boldsymbol{\mathcal  E})}_{n{\bf k}}|}.
\end{equation}
This allows the field-polarized states $\{u^{(\boldsymbol{\mathcal  E})}\}$  to be obtained using a direct optimization method
such as the conjugate gradient method~\cite{PhysRevB.55.10337,RevModPhys.64.1045}.
Following the work of Nunes and Gonze~\cite{PhysRevB.63.155107}, the gradient of the electronic polarization is given by
\begin{equation}\label{eq:deltaP}
\begin{split}
\frac{\delta ({\bf P}_{\rm el} \cdot \mathbf{b}_i) }{\delta \langle u^{(\boldsymbol{\mathcal  E})}_{\nk}|}
&  =
-\frac{\ii fe}{2\Omega N^{(i)}_{k^{(i)}_{\perp}} } \sum^N_{m=1}
\Bigg\{ \Big[M^{(\kk, \kk^{(i)}_{+})}\Big]^{-1}_{mn} | u^{(\boldsymbol{\mathcal  E})}_{m\mathbf{k}^{(i)}_{+}} \rangle  \\
& \quad \quad \quad \quad \quad \quad
-  \Big[M^{(\kk, \kk^{(i)}_{-})}\Big]^{-1}_{mn} | u^{(\boldsymbol{\mathcal  E})}_{m\mathbf{k}^{(i)}_{-}} \rangle \Bigg\},
\end{split}
\end{equation}
where $\mathbf{k}^{(i)}_{\pm} = \kk \pm \mathbf{b}_i / N^{(i)}_{k_{\parallel}}$.
Using Eq.~\eqref{eq:deltaP}, Eq.~\eqref{eq:gradient} becomes~\cite{PhysRevLett.89.117602}
\begin{equation}\label{eq:gradient_AE}
\begin{split}
\frac{\delta \mathcal{F}[\{u^{(\boldsymbol{\mathcal  E})}\},\boldsymbol{\mathcal  E}]}{\delta \langle u^{(\boldsymbol{\mathcal  E})}_{n\mathbf{k}}|}
&=
\frac{f}{N_\kk} H^{(\kk)}_{\rm KS} |u^{(\boldsymbol{\mathcal  E})}_{n{\bf k}}\rangle
+\frac{\ii fe}{4\pi } \sum_{i=1}^3  \frac{ \boldsymbol{\mathcal  E} \cdot \mathbf{a}_i }{N^{(i)}_{k^{(i)}_{\perp}}} \\
\times & \sum^N_{m=1}
\Bigg\{ \Big[M^{(\kk, \kk^{(i)}_{+})}\Big]^{-1}_{mn} | u^{(\boldsymbol{\mathcal  E})}_{m\mathbf{k}^{(i)}_{+}} \rangle -
 \Big[M^{(\kk, \kk^{(i)}_{-})}\Big]^{-1}_{mn} | u^{(\boldsymbol{\mathcal  E})}_{m\mathbf{k}^{(i)}_{-}} \rangle \Bigg\},
\end{split}
\end{equation}
where $\mathbf{a}_i$ is the real-space lattice vector satisfying $\mathbf{a}_i\cdot \mathbf{b}_j = 2\pi \delta_{ij}$.
In deriving Eq.~\eqref{eq:gradient_AE} one uses the fact that
\begin{equation}\label{eq:fact}
\boldsymbol{\mathcal  E} \cdot {\bf P}_{\rm el}
= \frac{1}{2\pi} \sum_{i=1}^3 (\boldsymbol{\mathcal  E} \cdot \mathbf{a}_i) ({\bf P}_{\rm el} \cdot \mathbf{b}_i).
\end{equation}

Finally, forces under a finite-electric-field can be calculated from the derivative of the electric enthalpy functional
\begin{equation}\label{eq:forces_general}
\begin{split}
\mathbf{F}^{(\boldsymbol{\mathcal  E})} \equiv
- \frac{d \mathcal{F}}{d \RR}
=  - \frac{\partial E^{(\boldsymbol{\mathcal  E})}_{\rm KS}}{\partial  \RR}
+ \Omega \boldsymbol{\mathcal  E} \cdot \frac{\partial {\bf P}_{\rm mac}}{\partial \RR},
\end{split}
\end{equation}
where the Hellmann-Feynman theorem is used.

\subsection{Electric field in the PAW formalism}\label{sec:EFIELD_PAW}

In this section, we summarize the expressions for the electronic polarization, the gradient of the electric enthalpy functional, and the
electric-field-induced forces within the PAW formalism.
Similar formulations can be found in Refs.~\cite{David1998_USPP, ZWANZIGER2012113}.

Within the PAW method, the all-electron orbitals $\psi_\nk$ are related to the
corresponding pseudo orbitals $\tilde\psi_\nk$ by a linear transformation~\cite{PhysRevB.50.17953,PhysRevB.59.1758}
\begin{equation} \label{eq:PAW_psi}
\ket{\psi_\nk }=\ket{\tilde\psi_\nk }+\sum_{\tau, \mu} \left(\ket{\phi_{\tau\mu}}-\ket{\tilde\phi_{\tau\mu}}\right)
\bra{\tilde{p}_{\tau\mu}}\tilde\psi_\nk\rangle.
\end{equation}
The  pseudo orbitals $\tilde\psi_\nk$ are the variational quantities of the PAW method and are expanded in plane waves.
$\phi_{\tau\mu}$ and $\tilde{\phi}_{\tau\mu}$ are the all-electron and pseudo partial waves, respectively, whereas $\tilde{p}_{\tau\mu}$ denotes
the corresponding projector functions, which are dual to
$\tilde{\phi}_{\tau\mu}$ within the PAW augmentation sphere ($\Omega_{\rm PAW}$).
The index $\tau$ represents the atomic site $\mathbf{R}_\tau$
and $\mu=(n_\mu, l_\mu, m_\mu)$ is an abbreviation for
the energy quantum number $n_\mu$ and
angular-momentum quantum numbers $(l_\mu,m_\mu)$ that characterize the
solution of the Schr\"odinger equation for a reference atom.
Within the PAW framework, the KS equations are generalized to~\cite{PhysRevB.59.1758}
\begin{equation} \label{eq:PAW_Ham}
\tilde{H}^{(\kk)}_{\rm KS} |\tilde{\psi}_{n{\bf k}}\rangle  = \epsilon_\nk \tilde{S}^{(\kk)}  |\tilde{\psi}_{n{\bf k}}\rangle,
\end{equation}
where $\tilde{H}^{(\kk)}_{\rm KS}$  and $\tilde{S}^{(\kk)}$  are the PAW Hamiltonian and overlap operator,
respectively.

The electronic polarization in the PAW approach can be split into two parts~\cite{David1998_USPP}:
\begin{equation}\label{Polarization_PAW}
{\bf P}_{\rm el}={\bf P}_{\rm EV}+{\bf P}_{\rm BP},
\end{equation}
where the first  ``expectation-value" (EV) term arises from the PAW charge augmentation, which is given by~\cite{David1998_USPP}
\begin{equation}\label{PEV_PAW}
{\bf P}_{\rm EV} \cdot {\bf b}_i =-\frac{fe}{\Omega N_{\kk}}
\sum\limits_{\tau,\mu,\mu',n,{\bf k}} ({\bf d}_{\tau\mu\mu'} \cdot {\bf b}_i)
\langle \tilde{u}_{n{\bf k}}|\tilde{p}^{({\bf k})}_{\tau\mu}\rangle \langle\tilde{p}^{({\bf k})}_{\tau\mu'}|\tilde{u}_{n{\bf k}}\rangle,
\end{equation}
where  $\tilde{u}_{n{\bf k}}$ is the cell-periodic part of $\tilde{\psi}_{n{\bf k}}$.
For brevity, we henceforth drop the field-polarized notation ${(\boldsymbol{\mathcal  E})}$ in the orbitals.
$|\tilde{p}^{({\bf k})}_{\tau\mu}\rangle \equiv e^{-i \kk \cdot (\rr-\RR_\tau)} |\tilde{p}_{\tau\mu}\rangle$,
and ${\bf d}_{\tau\mu\mu'}$  is the dipole moment of the one-center charge density
\begin{equation}\label{AugChgDipMoment_PAW}
{\bf d}_{\tau\mu\mu'}=\int_{\Omega_{\rm PAW}} \rr\, Q_{\tau\mu\mu'}\left({\bf r}\right)d\rr,
\end{equation}
with the one-center charge density defined by
\begin{equation}
Q_{\tau\mu\mu'}\left({\bf r}\right) = \phi^*_{\tau\mu}(\rr) \phi_{\tau\mu'}(\rr) -  \tilde{\phi}^*_{\tau\mu}(\rr) \tilde{\phi}_{\tau\mu'}(\rr).
\end{equation}
The second term in Eq.~\eqref{Polarization_PAW} is the so-called ``Berry-phase"  (BP) term~\cite{David1998_USPP}, which has the same form as
Eqs.~\eqref{eq:P_BP_1} and~\eqref{eq:P_BP_2}, except that
the overlap matrix [compare Eq.~\eqref{eq:overlap_all_electron}] should be adapted to the PAW method as
\begin{equation}\label{P_BP_PAW2}
\begin{split}
M^{(\kk_j, \kk_{j+1})}_{mn}
& = \langle \tilde{u}_{m{\bf k}_{j}}|\tilde{u}_{n{\bf k}_{j+1}}\rangle \\
& + \sum\limits_{\tau,\mu,\mu'}Q^{(\kk_j, \kk_{j+1})}_{\tau\mu\mu'}
\langle \tilde{u}_{m{\bf k}_{j}}|\tilde{p}^{({\bf k}_{j})}_{\tau\mu}\rangle \langle\tilde{p}^{({\bf k}_{j+1})}_{\tau\mu'} | \tilde{u}_{n{\bf k}_{j+1}}\rangle,
\end{split}
\end{equation}
where the norm of the cell periodic part of the one-center charge density is given by
\begin{equation}\label{eq:norm_onecenter}
Q^{(\kk_j, \kk_{j+1})}_{\tau\mu\mu'}=\int_{\Omega_{\rm PAW}} Q_{\tau\mu\mu'}\left({\bf r}\right) e^{-i (\kk_{j+1}-\kk_{j})\cdot \rr } d\rr.
\end{equation}

\begin{table*}
\caption{
PAW potentials used in the present calculations.
The exact VASP POTCAR identifiers, exchange-correlation (XC) functionals,
valence configurations, angular-momentum channels and their cutoff radii
$r_c$, outermost PAW cutoff radii $R_{\rm core}$, recommended plane-wave
energy cutoffs (ENMAX/ENMIN), and augmentation-charge cutoffs (EAUG) are listed.
The cutoff radii are given in Bohr and the energies in eV.
}
\begin{ruledtabular}
\begin{tabular}{ccccccccc}
System
& Element
& POTCAR identifier
& XC
& Valence configuration
& Channels: $r_c$
& $R_{\rm core}$
& ENMAX/ENMIN
& EAUG
\\
\hline
\multirow{4}{*}{AlX}
& Al
& PAW Al 17Apr2000
& LDA
& $3s^2 3p^1$
& $s$:1.90, $p$:1.90, $d$:1.90
& 1.90
& 240.957/180.718
& 291.052
\\
& P
& PAW P 26Mar2009
& LDA
& $3s^2 3p^3$
& $s$:1.90, $p$:1.90, $d$:1.90
& 1.90
& 255.155/191.366
& 342.924
\\
& As
& PAW As 15Mar2012
& LDA
& $4s^2 4p^3$
& $s$:2.10, $p$:2.10, $d$:2.10, $f$:2.10
& 2.10
& 208.870/156.653
& 462.758
\\
& Sb
& PAW Sb 04Feb1998
& LDA
& $5s^2 5p^3$
& $s$:2.30, $p$:2.30, $d$:2.30, $f$:2.30
& 2.30
& 172.301/129.225
& 380.513
\\
\hline
\multirow{3}{*}{AgNbO$_3$}
& Ag
& PAW\_PBE Ag 02Apr2005
& PBE
& $4d^{10}5s^1$
& $s$:2.40, $p$:2.50, $d$:2.40, $f$:2.40
& 2.50
& 249.844/187.383
& 412.496
\\
& Nb
& PAW\_PBE Nb\_pv 08Apr2002
& PBE
& $4p^6 4d^4 5s^1$
& $s$:2.75, $p$:2.60, $d$:2.60
& 2.75
& 208.608/156.456
& 355.067
\\
& O
& PAW\_PBE O 08Apr2002
& PBE
& $2s^2 2p^4$
& $s$:1.20, $p$:1.52, $d$:1.50
& 1.52
& 400.000/300.000
& 605.392
\\
\end{tabular}
\end{ruledtabular}
\label{table:PAW_potentials}
\end{table*}

The gradient of the electric enthalpy functional
[compare Eq.~\eqref{eq:gradient_AE}] in the PAW formalism can thus be derived as~\cite{ZWANZIGER2012113}
\begin{equation}
\begin{split}
& \frac{\delta \mathcal{F}[\{\tilde{u}\},\boldsymbol{\mathcal  E}]}{\delta \langle \tilde{u}_{n\mathbf{k}}|}
= \frac{f}{N_\kk} [\tilde{H}^{(\kk)}_{\rm KS} - \epsilon_\nk \tilde{S}^{(\kk)}]  |\tilde{u}_{n{\bf k}}\rangle \\
&\quad - \frac{fe}{2\pi N_\kk} \sum_{i=1}^3  (\boldsymbol{\mathcal  E} \cdot \mathbf{a}_i)
\sum\limits_{\tau,\mu,\mu'} ({\bf d}_{\tau\mu\mu'} \cdot {\bf b}_i) |\tilde{p}^{({\bf k})}_{\tau\mu}\rangle \langle\tilde{p}^{({\bf k})}_{\tau\mu'}|\tilde{u}_{n{\bf k}}\rangle \\
&\quad +\frac{\ii fe}{4\pi } \sum_{i=1}^3  \frac{(\boldsymbol{\mathcal  E} \cdot \mathbf{a}_i)}{N^{(i)}_{k^{(i)}_{\perp}} }
\sum^N_{m=1}
\Bigg\{ \Big[M^{(\kk, \kk^{(i)}_{+})}\Big]^{-1}_{mn} | \tilde{\xi}_{m\mathbf{k}^{(i)}_{+}} \rangle -
 \Big[M^{(\kk, \kk^{(i)}_{-})}\Big]^{-1}_{mn} | \tilde{\xi}_{m\mathbf{k}^{(i)}_{-}} \rangle \Bigg\}.
\end{split}
\end{equation}
Here, we have defined
\begin{equation}
\begin{split}
 | \tilde{\xi}_{m\mathbf{k}^{(i)}_\sigma} \rangle = | \tilde{u}_{m\kk_\sigma^{(i)}}\rangle + \sum\limits_{\tau,\mu,\mu'}Q^{(\kk, \kk_\sigma^{(i)})}_{\tau\mu\mu'}  |\tilde{p}^{(\kk)}_{\tau\mu}\rangle \langle\tilde{p}^{(\kk_\sigma^{(i)})}_{\tau\mu'}  | \tilde{u}_{m\kk_\sigma^{(i)}}\rangle,
\end{split}
\end{equation}
where $\sigma=+$ or $-$.

For the calculation of forces within the PAW formalism,  we note that
the explicit form of the first term $-{\partial E^{(\boldsymbol{\mathcal  E})}_{\rm KS}}/{\partial  \RR}$ in Eq.~\eqref{eq:forces_general}
has been given in Ref.~\cite{PhysRevB.59.1758} in detail except that here the field-polarized
orbitals must be used. The macroscopic polarization contribution [second term of Eq.~\eqref{eq:forces_general}]
to the forces is given by
\begin{equation}
\Omega \boldsymbol{\mathcal  E} \cdot \frac{\partial {\bf P}_{\rm mac}}{\partial \RR} =
\boldsymbol{\mathcal  E} Z^{\rm ion}_j e
+ \Omega \boldsymbol{\mathcal  E} \cdot \frac{\partial \mathbf{P}_{\rm el}}{\partial \RR},
\end{equation}
where the last term (electronic polarization contribution) in the PAW method contains two parts:
One is the on-site dipole contribution~\cite{ZWANZIGER2012113}
\begin{equation}\label{Force_EFIELD_dipole}
-\frac{fe}{2\pi N_\kk} \sum_{i=1}^3  (\boldsymbol{\mathcal  E} \cdot \mathbf{a}_i) \sum_{\kk,n}
\sum\limits_{\tau,\mu,\mu'} ({\bf d}_{\tau\mu\mu'} \cdot {\bf b}_i)
 \frac{\partial\Big(\langle \tilde u_\nk |\tilde{p}^{({\bf k})}_{\tau\mu}\rangle \langle\tilde{p}^{({\bf k})}_{\tau\mu'}|\tilde{u}_{n{\bf k}}\rangle\Big)}{\partial \RR},
\end{equation}
and the other one originates from the overlap matrix~\cite{ZWANZIGER2012113}
\begin{equation}\label{Force_EFIELD_overlap}
\begin{split}
& \frac{\ii fe}{4\pi } \sum_{i=1}^3  \frac{(\boldsymbol{\mathcal  E} \cdot \mathbf{a}_i)}{N^{(i)}_{k^{(i)}_{\perp}}}  \sum_{\kk,m,n}  \sum\limits_{\tau,\mu,\mu'} \\
 \Bigg\{
 & Q^{(\kk, \kk_{+}^{(i)})}_{\tau\mu\mu'}
\Big[M^{(\kk_i, \kk_{+}^{(i)})}\Big]^{-1}_{mn}
\frac{\partial \Big( \langle \tilde{u}_\nk |\tilde{p}^{(\kk)}_{\tau\mu}\rangle \langle\tilde{p}^{(\kk_{+}^{(i)})}_{\tau\mu'}  | \tilde{u}_{m{\kk_{+}^{(i)}}}\rangle \Big)}{\partial \RR} \\
-&Q^{(\kk, \kk_{-}^{(i)})}_{\tau\mu\mu'}
\Big[M^{(\kk_i, \kk_{-}^{(i)})}\Big]^{-1}_{mn}
\frac{\partial \Big( \langle \tilde{u}_\nk |\tilde{p}^{(\kk)}_{\tau\mu}\rangle \langle\tilde{p}^{(\kk_{-}^{(i)})}_{\tau\mu'}  | \tilde{u}_{m{\kk_{-}^{(i)}}}\rangle \Big)}{\partial \RR}
\Bigg\}. \\
\end{split}
\end{equation}

\section{Computational details}\label{sec:tech}

The finite-electric-field approach within the PAW formalism has been implemented in the VASP code~\cite{PhysRevB.47.558, PhysRevB.54.11169}.
In this work, the field-induced forces in Eq.~\eqref{eq:forces_general} are coupled to the ionic relaxation procedure. At each ionic step, VASP converges the electronic state at the applied field, evaluates the forces, and passes them to the ionic optimizer to update the atomic positions while maintaining the field.
The PAW datasets~\cite{PhysRevB.59.1758} used in the present calculations are summarized in
Table~\ref{table:PAW_potentials}, including their exact POTCAR identifiers,
valence configurations, angular-momentum channels, and characteristic cutoff parameters.
For As and Sb, we verified that including the $d$ electrons in the valence has a marginal effect on the computed properties,
such as Born effective charges, dielectric constants, and piezoelectric constants.
For the atomic relaxations under an electric field, the conjugate gradient algorithm~\cite{RevModPhys.64.1045} was used for the electronic optimizations.

For AlX (X=P, As, and Sb), the local density approximation (LDA)~\cite{PhysRevB.23.5048} in the parametrization of Ceperley and Alder~\cite{PhysRevLett.45.566}
was used. We employed a plane-wave cutoff of 500 eV and a $\Gamma$-centered 16$\times$16$\times$16 $k$-point grid,
which were previously shown to provide converged dielectric constant~\cite{ZWANZIGER2012113} and quadratic dielectric susceptibility~\cite{PhysRevB.71.125107}.
The electronic iterations were considered converged when the total-energy difference between successive iterations was below $10^{-7}$ eV.
The ionic relaxation was continued until the residual force on each atom was below $10^{-3}$ eV/\AA.

For the extraction of the response coefficients below, the calculated polarization and electric field were expressed in SI units.
Expanding the macroscopic polarization to second order in the electric field gives
\begin{equation}\label{eq:P_vs_E}
(P_{\rm mac})_\alpha = (P^0_{\rm mac})_\alpha +\epsilon_0 \sum_\beta \chi^{(1)}_{\alpha\beta}\mathcal{E}_\beta
+ \epsilon_0  \sum_{\beta\gamma} \chi^{(2)}_{\alpha\beta\gamma}\mathcal{E}_\beta  \mathcal{E}_\gamma.
\end{equation}
Here, $\alpha,\beta,\gamma=x,y,z$, $(P^0_{\rm mac})_\alpha$ is the zero-field polarization, and $\chi^{(1)}$ and $\chi^{(2)}$ are the linear and quadratic dielectric susceptibilities, respectively.
The relative dielectric tensor is then given by
\begin{equation}\label{eq:die}
\epsilon_{\alpha\beta}
=\delta_{\alpha\beta}+ \chi^{(1)}_{\alpha\beta}
=\delta_{\alpha\beta}+ \frac{1}{\epsilon_0}\frac{\partial (P_{\rm mac})_\alpha}{\partial {\mathcal  E}_\beta}.
\end{equation}
For the static response considered here, the intrinsic permutation symmetry
$\chi^{(2)}_{\alpha\beta\gamma} = \chi^{(2)}_{\alpha\gamma\beta}$ gives
\begin{equation}\label{eq:die2}
\chi^{(2)}_{\alpha\beta\gamma}=\frac{1}{2\epsilon_0} \frac{\partial^2  (P_{\rm mac})_\alpha}{\partial{\mathcal  E}_\beta \partial{\mathcal  E}_\gamma}.
\end{equation}
The Born effective charge tensor for the $k$-th ion was calculated by~\cite{PhysRevLett.89.117602}
\begin{equation}\label{eq:Born}
Z^*_{k,\alpha\beta} =\frac{1}{e}\frac{\partial F_{k,\beta}}{\partial {\mathcal  E}_\alpha},
\end{equation}
where $F_{k,\beta}$ is the force acting on the $k$-th ion along the $\beta$ coordinate.
The piezoelectric tensor was calculated by~\cite{PhysRevLett.89.117602,VANDERBILT2000147}
\begin{equation}\label{eq:pie}
c_{\alpha\beta\gamma}=-\frac{\partial \sigma_{\alpha\beta}}{\partial \mathcal{E}_\gamma}.
\end{equation}
For comparison with experiment, only the proper piezoelectric tensor was calculated, i.e.,
the polarization contribution to the stress was not included.
To validate the finite-electric-field implementation, the dielectric, piezoelectric, and Born effective charge tensors were also computed
by DFPT~\cite{PhysRevB.55.10337,PhysRevB.73.045112,RevModPhys.73.515,PhysRevB.72.035105} using the VASP code.

For AgNbO$_3$, the Perdew-Burke-Ernzerhof (PBE)~\cite{PhysRevLett.77.3865} approximation
was employed for the exchange-correlation functional.
The antiferroelectric $Pbcm$ phase was first fully relaxed (without electric field)
using a plane-wave cutoff of 600 eV and a $\Gamma$-centered 5$\times$5$\times$2  $k$-point grid.
The optimized lattice parameters were $a$=5.653 $\AA$, $b$=5.741 $\AA$, and $c$=15.760 $\AA$,
which, as expected, overestimate the experimental values $a$=5.544 $\AA$, $b$=5.607 $\AA$, and $c$=15.565 $\AA$~\cite{Sciau_2004}.
Then, the electric field was applied along the $\mathbf{b}$ axis and atomic positions
were relaxed using a reduced energy cutoff (544 eV) and $k$-point grid (3$\times$3$\times$2)
until the forces acting on each atom were less than 0.05 eV/$\AA$.
These settings were chosen to follow Ref.~\cite{doi:10.1063/1.4941319}, thereby allowing a
direct comparison with the previous calculations.

\section{Results}\label{sec:results}

\subsection{AlX (X=P, As, and Sb)}\label{sec:results_AlP}

\begin{table*}
\caption{Computed lattice parameters $a$, Born effective charges $Z^*$, ion-clamped ($\epsilon^\infty$) and unclamped ($\epsilon^0$)
dielectric constants, ion-clamped quadratic dielectric susceptibility $\chi^{(2)}_{xyz}$ (in pm/V), and
ion-clamped ($c^0_{xyz}$) and unclamped ($c_{xyz}$) proper piezoelectric constants (in $\mu$C/cm$^2$) for AlX (X=P, As, and Sb) semiconductors.
``Opt." and ``Expt." indicate that the calculations were conducted using the LDA optimized and experimental lattice parameters, respectively.
Available theoretical and experimental data are given for comparison.
Note that the lattice parameters used in Ref.~\cite{ZWANZIGER2012113} were not explicitly reported, but were obtained from LDA optimizations using PAW potentials.
}
\begin{ruledtabular}
\begin{tabular}{ccccccccc}
 &   &  $a$ (\AA) & $Z^*$    & $\epsilon^\infty$ & $\epsilon^0$ & $\chi^{(2)}_{xyz}$ &  $c^0_{xyz}$  & $c_{xyz}$  \\
  \hline
\multirow{6}{0.3cm}{AlP}
& Finite field (This work, Opt.)                 & 5.435  & 2.21 &   8.34 & 10.48 & 43  & $-67.77$ & 4.77       \\
& DFPT (This work, Opt.)                         & 5.435  & 2.21 &   8.34 & 10.49 & --- & $-67.72$ & 4.77     \\
& Finite field (This work, Expt.)                 & 5.463~\cite{Richman_1968}  & 2.22 &   8.36 & 10.58 & 44  & $-67.94$ & 0.90       \\
& Finite field~\cite{PhysRevLett.89.117602}& 5.419  & 2.24 &   8.1  & 10.2  & 39  & $-71.48$ & 2.73    \\
& Finite field~\cite{ZWANZIGER2012113}    & --- & 2.22 &   8.01 & 10.26 & 43  & --- & --- \\
& Experiments                                    & 5.463~\cite{Richman_1968}  & 2.28~\cite{PhysRevB.4.1367}  &   7.4~\cite{Adachi2005}  & 9.8~\cite{MONEMAR19701295}   & --- & --- & --- \\
 \hline
\multirow{6}{0.3cm}{AlAs}
& Finite field (This work, Opt.)                 & 5.638  & 2.13 &   9.45 & 11.40 & 74  & $-65.80$ & $-1.81$     \\
& DFPT (This work, Opt.)                         & 5.638  & 2.13 &   9.46 & 11.41 & --- & $-65.83$ & $-1.82$  \\
& Finite field (This work, Expt.)                 & 5.661~\cite{doi:10.1063/1.1658389}  & 2.14 &  9.53 & 11.52 & 79  & $-66.13$ & $-4.66$       \\
& Finite field~\cite{PhysRevLett.89.117602}& 5.604  & 2.14 &   9.6  & 11.5  & 64  & $-71.43 $ & $-5.10$  \\
& Finite field~\cite{ZWANZIGER2012113}     & --- & 2.18 &   8.75 & 11.05 & 65  & --- & --- \\
& Experiments                                    & 5.661~\cite{doi:10.1063/1.1658389}  & 2.20~\cite{PhysRevB.50.14125} &   8.16~\cite{GUYAUX1992697} & 10.06~\cite{Adachi2005} & --- & --- & --- \\
 \hline
\multirow{6}{0.3cm}{AlSb}
& Finite field (This work, Opt.)                 & 6.121  & 1.81 &  11.73 & 13.00 & 193 & $-52.29$ & $-4.95$     \\
& DFPT (This work, Opt.)                         & 6.121  & 1.81 &  11.77 & 13.04 & --- & $-52.25$ & $-4.94$  \\
& Finite field (This work, Expt.)                 & 6.136~\cite{Giesecke:a02256}  & 1.82 &   11.80 & 13.09 & 200  & $-52.61$ & $-6.18$       \\
& Finite field~\cite{ZWANZIGER2012113}     & --- & 1.84 &  11.17 & 12.54 & 197 & --- & --- \\
& Experiments                                   & 6.136~\cite{Giesecke:a02256}  & 1.93~\cite{PhysRevB.4.1367} &  10.24~\cite{HASS19621099} & 12.04~\cite{HASS19621099} & 153$\pm$13~\cite{LandoltBornstein2000:sm_lbs_978-3-540-37124-3_40} & --- & $-6.81$~\cite{PhysRevLett.62.2853}       \\
&                           &  & &   9.88~\cite{PhysRev.127.126} & & 196~\cite{Singh1986_book} & &       \\
\end{tabular}
\end{ruledtabular}
\label{table:AlX}
\end{table*}

\begin{figure}
\begin{center}
\includegraphics[width=0.50\textwidth,clip]{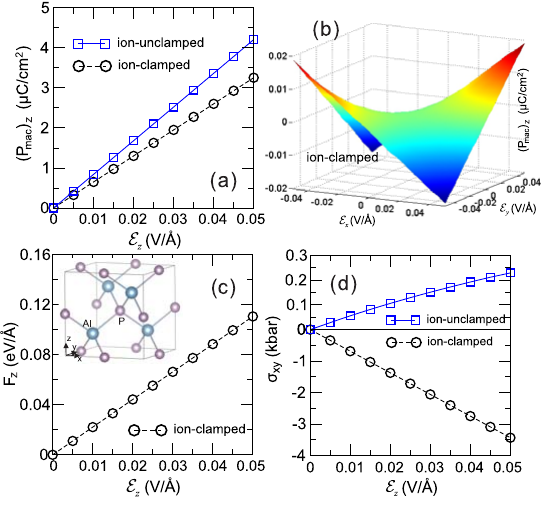}
\end{center}
 \caption{Illustration of (a) polarization $(P_{\rm mac})_z$ vs. $\mathcal{E}_z$ for ion-clamped (circles) and unclamped (squares) cases,
 (b) $(P_{\rm mac})_z$ vs. $\mathcal{E}_x$ and $\mathcal{E}_y$ for clamped ions,
 (c) the force $F_z$ on the Al atom  vs. $\mathcal{E}_z$, and (d) stress $\sigma_{xy}$ vs. $\mathcal{E}_z$ for AlP, whose
 crystal structure is given in the inset of (c).
}
\label{fig:AlP}
\end{figure}

AlX (X=P, As, and Sb) semiconductors crystallize in the zinc-blende structure (space group $F\bar{4}3m$) [see the inset of Fig.~\ref{fig:AlP}(c)],
which is non-centrosymmetric but has no spontaneous polarization.
Figure~\ref{fig:AlP} shows the calculated polarization, force, and stress as functions of the applied electric field for AlP as a representative example.
As expected, at small electric fields, the electronic polarization increases linearly with the electric field.
Compared to the electronic polarization, the ionic contribution for AlP is small [Fig.~\ref{fig:AlP}(a)].
As the electric field increases, the nonlinear behavior becomes apparent [Fig.~\ref{fig:AlP}(b)].
The dielectric constants can be calculated from Eq.~\eqref{eq:die}
and quadratic dielectric susceptibilities can be obtained from Eq.~\eqref{eq:die2}
by using the finite difference method around the zero field.
Similarly, the Born effective charge and piezoelectric constant, which are the force and stress responses to the electric field, respectively,
can be obtained from Eq.~\eqref{eq:Born} and Eq.~\eqref{eq:pie}.
We note that for AlX (X=P, As, and Sb), by symmetry, there is only one nonzero independent
dielectric constant ($\epsilon_{xx}=\epsilon_{yy}=\epsilon_{zz}$),
Born effective charge ($Z^*_{xx}=Z^*_{yy}=Z^*_{zz}$), quadratic dielectric susceptibility  ($\chi^{(2)}_{xyz}=\chi^{(2)}_{yzx}=\chi^{(2)}_{zxy}$)
as well as piezoelectric constant ($c_{xyz}=c_{yzx}=c_{zxy}$).

Table~\ref{table:AlX} summarizes the calculated Born effective charges, ion-clamped and unclamped dielectric constants,
ion-clamped quadratic dielectric susceptibility, and ion-clamped and unclamped proper piezoelectric constants for this series of semiconductors AlX (X=P, As, and Sb).
One can observe that the finite-electric-field results agree well with DFPT for the Born effective charges, linear dielectric constants, and piezoelectric coefficients.
Note that the implementation of quadratic dielectric susceptibilities within DFPT is technically more involved because they correspond to a higher-order
response and are commonly evaluated using the $2n+1$ theorem.
In contrast, within the finite-field approach they can be obtained directly from finite differences of the polarization with respect to the applied field.
The DFPT values obtained using the $2n+1$ theorem are 39, 64, and 146 pm/V for AlP, AlAs, and AlSb, respectively~\cite{PhysRevB.71.125107,PhysRevB.53.15638},
compared with our finite-field values of 43, 74, and 193 pm/V.
The corresponding differences are approximately 10\%, 16\%, and 32\%, respectively; thus, the AlSb value shows a noticeable quantitative discrepancy from DFPT.
By contrast, our results are generally consistent with other theoretical data obtained using the same finite-electric-field method~\cite{PhysRevLett.89.117602, ZWANZIGER2012113}.
The small deviations might arise from the different lattice parameters caused by different pseudopotentials and different implementation details and technical settings used.
As expected from the additional contribution of internal ionic relaxations,
the ion-unclamped piezoelectric constants exhibit a much stronger dependence on the lattice parameters (see Table~\ref{table:AlX}).
Finally, except for the overestimation of the dielectric constants by LDA due to its underestimation of the band gap,
the overall agreement between our calculated results and experiments is satisfactory.
Although there are no experimental data on ion-unclamped  proper piezoelectric constants $c_{xyz}$ for AlX (X=P and As),
the calculated results follow the chemical trends across the series, i.e., $c_{xyz}$ increases as X goes from Sb via As to P.
Due to the nearly canceling contributions from electrons and ions, it is thus unsurprising that
$c_{xyz}$ becomes positive for AlP  [see Fig.~\ref{fig:AlP}(d)].

In summary, the Born effective charges, dielectric constants, quadratic dielectric susceptibilities, and piezoelectric constants for AlX (X=P and As)
calculated using the electric-field approach are in good overall agreement with available experimental and theoretical data, confirming the reliability of this approach within the PAW formalism.

\subsection{AgNbO$_3$}\label{sec:results_AgNbO3}

\begin{figure}[bp]
\begin{center}
\includegraphics[width=0.48\textwidth,clip]{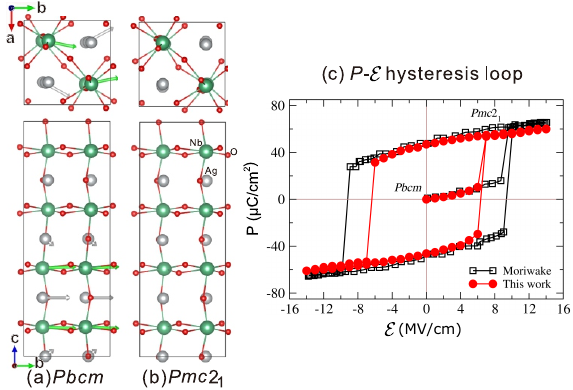}
\end{center}
 \caption{Finite electric-field-induced structural phase transition of AgNbO$_3$ from (a) the antiferroelectric $Pbcm$ phase to
 (b) the ferroelectric $Pmc2_1$ phase, calculated using a 40-atom unit cell.
 Displacements of Nb and Ag ions driven by the electric field along the $\mathbf{b}$ axis
 are indicated as arrows in (a). Small displacements of O ions are not shown for clarity.
 (c) Computed $P$-$\mathcal{E}$  hysteresis loop,
 showing good agreement with the calculated results of Moriwake~\cite{doi:10.1063/1.4941319}.
}
\label{fig:AgNbO3}
\end{figure}

Now we turn to the more complex case of perovskite-structured AgNbO$_3$.
This compound---as a candidate parent material for lead-free piezoelectric applications---attracts considerable interest
because it undergoes a series of complicated phase transitions from
paraelectric cubic ($Pm\bar{3}m$), to paraelectric tetragonal ($P4/mbm$),
paraelectric orthorhombic ($Cmcm$), antiferroelectric orthorhombic ($Pbcm$), and ferroelectric orthorhombic ($Pmc2_1$)
as the temperature decreases~\cite{Sciau_2004, PhysRevB.79.104113, Yashima_2011,doi:10.1111/j.1551-2916.2012.05392.x,TianEtAl2022AgNbO3}.
Therefore, understanding the mechanism of the phase transitions is essential, in particular from $Pbcm$ to $Pmc2_1$.
First-principles calculations showed that the antiferroelectric phase is dynamically stable~\cite{Moriwake_2012},
thus excluding a soft-phonon-driven mechanism for the phase transition. In addition, it was found that the antiferroelectric phase
was slightly lower in energy by only 0.5 meV/f.u. than the ferroelectric phase
and the potential barrier required to overcome was small (40 meV/f.u),
suggesting the coexistence of both phases~\cite{doi:10.1063/1.4941319}. This is consistent with experimental observations that
the polarization-electric field hysteresis loop exhibits antiferroelectric behavior but with a weak ferroelectricity
at zero field~\cite{doi:10.1063/1.2751136,doi:10.1111/jace.15780}.

To explain the antiferroelectric-to-ferroelectric phase transition of AgNbO$_3$,
Moriwake and co-workers demonstrated using first-principles calculations that this transition can be driven by an applied electric field~\cite{doi:10.1063/1.4941319}.
In their calculations, electric-field-induced forces were used to perform structural relaxations using the ABINIT code~\cite{GONZE2002478,PhysRevLett.89.117602,ZWANZIGER2012113}.
For this reason, we chose AgNbO$_3$ as our second case to validate the finite-electric-field implementation.

Figure~\ref{fig:AgNbO3} shows our computed polarization-electric field hysteresis loop of AgNbO$_3$, highlighting
the structural phase transition from the antiferroelectric $Pbcm$ to the ferroelectric $Pmc2_1$ phase.
First, we observe that at small electric fields, the polarization characterized by the Nb and Ag ionic displacements indicated in Fig.~\ref{fig:AgNbO3}(a)
increases almost linearly with the electric field. As the electric field is increased to the critical value of 7 MV/cm,
the polarization abruptly increases to a large value of 53.8 $\mu$C/cm$^2$, indicating the phase transition
to the $Pmc2_1$ phase  [Fig.~\ref{fig:AgNbO3}(b)]. Our calculated polarization at the critical field is in good agreement with the experimental value of
52 $\mu$C/cm$^2$~\cite{doi:10.1063/1.2751136}. Interestingly, when the electric field is removed, the obtained
ferroelectric $Pmc2_1$ phase does not relax back to the original antiferroelectric $Pbcm$ phase, but remains in the ferroelectric state
with a slightly reduced polarization of 46.0 $\mu$C/cm$^2$.
This is due to the existence of an energy barrier of 40 meV/f.u.~\cite{doi:10.1063/1.4941319}.
Second, our calculated polarization-electric field hysteresis loop
agrees well with the calculated result from Moriwake~\cite{doi:10.1063/1.4941319} except that in our case we obtained a lower critical
field for the phase transition [7 MV/cm (this work) vs. 9.6 MV/cm~\cite{doi:10.1063/1.4941319}].
Sensitivity tests varying the plane-wave cutoff, $k$-point mesh, and force-convergence threshold
showed that these settings did not materially change the calculated transition field.
This difference is partly attributable to the different optimized cell volumes obtained with the different pseudopotentials [PAW (this work) vs. norm-conserving~\cite{doi:10.1063/1.4941319}],
which yield unit-cell volumes of 511.5 $\AA^3$ (this work) and 503.2 $\AA^3$~\cite{doi:10.1063/1.4941319}.
The larger volume in our case results in a lower critical electric field for the phase transition.
This was confirmed by our test calculations:
if the optimized lattice parameters of the $Pbcm$ phase from Moriwake~\cite{note} were used, a relatively larger transition electric field (8 MV/cm) was obtained.
The remaining difference may arise from the different finite-field implementations, although the exact origin remains unclear to us.
Thus, despite the remaining quantitative difference in the critical field, the reproduction of the field-induced antiferroelectric-to-ferroelectric transition and the overall agreement in the $P$--$\mathcal{E}$ hysteresis behavior between VASP and ABINIT provide a cross-code validation of the electric-field-induced-force implementation in VASP.

\section{Conclusions}\label{sec:conclusions}
To summarize, we have assessed the calculation of finite-electric-field-induced forces for bulk solids within the PAW formalism.
This capability enables the calculation of linear and nonlinear responses to electric fields and allows \emph{ab initio} structural relaxations of semiconductors and insulators under external electric fields. The reliability of this approach has been validated through computed ion-clamped and unclamped dielectric and piezoelectric constants, quadratic dielectric susceptibilities, and Born effective charges for semiconductors AlX (X = P, As, Sb), as well as through the reproduction of the electric-field-driven antiferroelectric-to-ferroelectric phase transition in AgNbO$_3$.
All results are in good agreement with available experimental and theoretical data.
We anticipate that this capability, as implemented in the VASP code, will facilitate future \emph{ab initio} investigations of electric-field effects in condensed matter,
including electric-field-controlled structural distortions, magnetizations, shift currents, Rashba effects, heat fluxes, and related phenomena.

\begin{acknowledgments}
This work was supported by the National Natural Science Foundation of China (Grant No.~52422112)
and the Austrian Science Fund (FWF) within the  SFB TACO  (Project No. 10.55776/F81).
P. Liu thanks J. He,  H. Zhang, and H. Moriwake for useful discussions.
\end{acknowledgments}

\bibliographystyle{apsrev4-1}
\bibliography{reference} 

\begin{thebibliography}{86}%
\makeatletter
\providecommand \@ifxundefined [1]{%
 \@ifx{#1\undefined}
}%
\providecommand \@ifnum [1]{%
 \ifnum #1\expandafter \@firstoftwo
 \else \expandafter \@secondoftwo
 \fi
}%
\providecommand \@ifx [1]{%
 \ifx #1\expandafter \@firstoftwo
 \else \expandafter \@secondoftwo
 \fi
}%
\providecommand \natexlab [1]{#1}%
\providecommand \enquote  [1]{``#1''}%
\providecommand \bibnamefont  [1]{#1}%
\providecommand \bibfnamefont [1]{#1}%
\providecommand \citenamefont [1]{#1}%
\providecommand \href@noop [0]{\@secondoftwo}%
\providecommand \href [0]{\begingroup \@sanitize@url \@href}%
\providecommand \@href[1]{\@@startlink{#1}\@@href}%
\providecommand \@@href[1]{\endgroup#1\@@endlink}%
\providecommand \@sanitize@url [0]{\catcode `\\12\catcode `\$12\catcode
  `\&12\catcode `\#12\catcode `\^12\catcode `\_12\catcode `\%12\relax}%
\providecommand \@@startlink[1]{}%
\providecommand \@@endlink[0]{}%
\providecommand \url  [0]{\begingroup\@sanitize@url \@url }%
\providecommand \@url [1]{\endgroup\@href {#1}{\urlprefix }}%
\providecommand \urlprefix  [0]{URL }%
\providecommand \Eprint [0]{\href }%
\providecommand \doibase [0]{http://dx.doi.org/}%
\providecommand \selectlanguage [0]{\@gobble}%
\providecommand \bibinfo  [0]{\@secondoftwo}%
\providecommand \bibfield  [0]{\@secondoftwo}%
\providecommand \translation [1]{[#1]}%
\providecommand \BibitemOpen [0]{}%
\providecommand \bibitemStop [0]{}%
\providecommand \bibitemNoStop [0]{.\EOS\space}%
\providecommand \EOS [0]{\spacefactor3000\relax}%
\providecommand \BibitemShut  [1]{\csname bibitem#1\endcsname}%
\let\auto@bib@innerbib\@empty
\bibitem [{\citenamefont {Ohno}\ \emph {et~al.}(2000)\citenamefont {Ohno},
  \citenamefont {Chiba}, \citenamefont {Matsukura}, \citenamefont {Omiya},
  \citenamefont {Abe}, \citenamefont {Dietl}, \citenamefont {Ohno},\ and\
  \citenamefont {Ohtani}}]{Ohno_Nature2000}%
  \BibitemOpen
  \bibfield  {author} {\bibinfo {author} {\bibfnamefont {H.}~\bibnamefont
  {Ohno}}, \bibinfo {author} {\bibfnamefont {D.}~\bibnamefont {Chiba}},
  \bibinfo {author} {\bibfnamefont {F.}~\bibnamefont {Matsukura}}, \bibinfo
  {author} {\bibfnamefont {T.}~\bibnamefont {Omiya}}, \bibinfo {author}
  {\bibfnamefont {E.}~\bibnamefont {Abe}}, \bibinfo {author} {\bibfnamefont
  {T.}~\bibnamefont {Dietl}}, \bibinfo {author} {\bibfnamefont
  {Y.}~\bibnamefont {Ohno}}, \ and\ \bibinfo {author} {\bibfnamefont
  {K.}~\bibnamefont {Ohtani}},\ }\href {\doibase
  https://doi.org/10.1038/35050040} {\bibfield  {journal} {\bibinfo  {journal}
  {Nature}\ }\textbf {\bibinfo {volume} {408}},\ \bibinfo {pages} {944}
  (\bibinfo {year} {2000})}\BibitemShut {NoStop}%
\bibitem [{\citenamefont {Zhao}\ \emph {et~al.}(2006)\citenamefont {Zhao},
  \citenamefont {Scholl}, \citenamefont {Zavaliche}, \citenamefont {Lee},
  \citenamefont {Barry}, \citenamefont {Doran}, \citenamefont {Cruz},
  \citenamefont {Chu}, \citenamefont {Ederer}, \citenamefont {Spaldin},
  \citenamefont {Das}, \citenamefont {Kim}, \citenamefont {Baek}, \citenamefont
  {Eom},\ and\ \citenamefont {Ramesh}}]{Zhao_NM_2006}%
  \BibitemOpen
  \bibfield  {author} {\bibinfo {author} {\bibfnamefont {T.}~\bibnamefont
  {Zhao}}, \bibinfo {author} {\bibfnamefont {A.}~\bibnamefont {Scholl}},
  \bibinfo {author} {\bibfnamefont {F.}~\bibnamefont {Zavaliche}}, \bibinfo
  {author} {\bibfnamefont {K.}~\bibnamefont {Lee}}, \bibinfo {author}
  {\bibfnamefont {M.}~\bibnamefont {Barry}}, \bibinfo {author} {\bibfnamefont
  {A.}~\bibnamefont {Doran}}, \bibinfo {author} {\bibfnamefont {M.~P.}\
  \bibnamefont {Cruz}}, \bibinfo {author} {\bibfnamefont {Y.~H.}\ \bibnamefont
  {Chu}}, \bibinfo {author} {\bibfnamefont {C.}~\bibnamefont {Ederer}},
  \bibinfo {author} {\bibfnamefont {N.~A.}\ \bibnamefont {Spaldin}}, \bibinfo
  {author} {\bibfnamefont {R.~R.}\ \bibnamefont {Das}}, \bibinfo {author}
  {\bibfnamefont {D.~M.}\ \bibnamefont {Kim}}, \bibinfo {author} {\bibfnamefont
  {S.~H.}\ \bibnamefont {Baek}}, \bibinfo {author} {\bibfnamefont {C.~B.}\
  \bibnamefont {Eom}}, \ and\ \bibinfo {author} {\bibfnamefont
  {R.}~\bibnamefont {Ramesh}},\ }\href {\doibase
  https://doi.org/10.1038/nmat1731} {\bibfield  {journal} {\bibinfo  {journal}
  {Nature materials}\ }\textbf {\bibinfo {volume} {5}},\ \bibinfo {pages} {823}
  (\bibinfo {year} {2006})}\BibitemShut {NoStop}%
\bibitem [{\citenamefont {Chiba}\ \emph {et~al.}(2008)\citenamefont {Chiba},
  \citenamefont {Sawicki}, \citenamefont {Nishitani}, \citenamefont {Nakatani},
  \citenamefont {Matsukura},\ and\ \citenamefont {Ohno}}]{Chiba_Nature2008}%
  \BibitemOpen
  \bibfield  {author} {\bibinfo {author} {\bibfnamefont {D.}~\bibnamefont
  {Chiba}}, \bibinfo {author} {\bibfnamefont {M.}~\bibnamefont {Sawicki}},
  \bibinfo {author} {\bibfnamefont {Y.}~\bibnamefont {Nishitani}}, \bibinfo
  {author} {\bibfnamefont {Y.}~\bibnamefont {Nakatani}}, \bibinfo {author}
  {\bibfnamefont {F.}~\bibnamefont {Matsukura}}, \ and\ \bibinfo {author}
  {\bibfnamefont {H.}~\bibnamefont {Ohno}},\ }\href {\doibase
  https://doi.org/10.1038/nature07318} {\bibfield  {journal} {\bibinfo
  {journal} {Nature}\ }\textbf {\bibinfo {volume} {455}},\ \bibinfo {pages}
  {515} (\bibinfo {year} {2008})}\BibitemShut {NoStop}%
\bibitem [{\citenamefont {Heron}\ \emph
  {et~al.}(2014{\natexlab{a}})\citenamefont {Heron}, \citenamefont {Bosse},
  \citenamefont {He}, \citenamefont {Gao}, \citenamefont {Trassin},
  \citenamefont {Ye}, \citenamefont {Clarkson}, \citenamefont {Wang},
  \citenamefont {Liu}, \citenamefont {Salahuddin}, \citenamefont {Ralph},
  \citenamefont {Schlom}, \citenamefont {\'I\~niguez}, \citenamefont {Huey},\
  and\ \citenamefont {Ramesh}}]{Heron_Nature_2014}%
  \BibitemOpen
  \bibfield  {author} {\bibinfo {author} {\bibfnamefont {J.~T.}\ \bibnamefont
  {Heron}}, \bibinfo {author} {\bibfnamefont {J.~L.}\ \bibnamefont {Bosse}},
  \bibinfo {author} {\bibfnamefont {Q.}~\bibnamefont {He}}, \bibinfo {author}
  {\bibfnamefont {Y.}~\bibnamefont {Gao}}, \bibinfo {author} {\bibfnamefont
  {M.}~\bibnamefont {Trassin}}, \bibinfo {author} {\bibfnamefont
  {L.}~\bibnamefont {Ye}}, \bibinfo {author} {\bibfnamefont {J.~D.}\
  \bibnamefont {Clarkson}}, \bibinfo {author} {\bibfnamefont {C.}~\bibnamefont
  {Wang}}, \bibinfo {author} {\bibfnamefont {J.}~\bibnamefont {Liu}}, \bibinfo
  {author} {\bibfnamefont {S.}~\bibnamefont {Salahuddin}}, \bibinfo {author}
  {\bibfnamefont {D.~C.}\ \bibnamefont {Ralph}}, \bibinfo {author}
  {\bibfnamefont {D.~G.}\ \bibnamefont {Schlom}}, \bibinfo {author}
  {\bibfnamefont {J.}~\bibnamefont {\'I\~niguez}}, \bibinfo {author}
  {\bibfnamefont {B.~D.}\ \bibnamefont {Huey}}, \ and\ \bibinfo {author}
  {\bibfnamefont {R.}~\bibnamefont {Ramesh}},\ }\href {\doibase
  https://doi.org/10.1038/nature14004} {\bibfield  {journal} {\bibinfo
  {journal} {Nature}\ }\textbf {\bibinfo {volume} {516}},\ \bibinfo {pages}
  {370} (\bibinfo {year} {2014}{\natexlab{a}})}\BibitemShut {NoStop}%
\bibitem [{\citenamefont {Gao}\ \emph {et~al.}(2017)\citenamefont {Gao},
  \citenamefont {Zhang}, \citenamefont {Ratcliff}, \citenamefont {Maruyama},
  \citenamefont {Murakami}, \citenamefont {Varatharajan}, \citenamefont
  {Yamani}, \citenamefont {Chen}, \citenamefont {Wang}, \citenamefont {Zhang},
  \citenamefont {Shull}, \citenamefont {Bendersky}, \citenamefont {Unguris},
  \citenamefont {Ramesh},\ and\ \citenamefont {Takeuchi}}]{Gao_NanoLett2017}%
  \BibitemOpen
  \bibfield  {author} {\bibinfo {author} {\bibfnamefont {T.}~\bibnamefont
  {Gao}}, \bibinfo {author} {\bibfnamefont {X.}~\bibnamefont {Zhang}}, \bibinfo
  {author} {\bibfnamefont {W.}~\bibnamefont {Ratcliff}}, \bibinfo {author}
  {\bibfnamefont {S.}~\bibnamefont {Maruyama}}, \bibinfo {author}
  {\bibfnamefont {M.}~\bibnamefont {Murakami}}, \bibinfo {author}
  {\bibfnamefont {A.}~\bibnamefont {Varatharajan}}, \bibinfo {author}
  {\bibfnamefont {Z.}~\bibnamefont {Yamani}}, \bibinfo {author} {\bibfnamefont
  {P.}~\bibnamefont {Chen}}, \bibinfo {author} {\bibfnamefont {K.}~\bibnamefont
  {Wang}}, \bibinfo {author} {\bibfnamefont {H.}~\bibnamefont {Zhang}},
  \bibinfo {author} {\bibfnamefont {R.}~\bibnamefont {Shull}}, \bibinfo
  {author} {\bibfnamefont {L.~A.}\ \bibnamefont {Bendersky}}, \bibinfo {author}
  {\bibfnamefont {J.}~\bibnamefont {Unguris}}, \bibinfo {author} {\bibfnamefont
  {R.}~\bibnamefont {Ramesh}}, \ and\ \bibinfo {author} {\bibfnamefont
  {I.}~\bibnamefont {Takeuchi}},\ }\href
  {https://doi.org/10.1021/acs.nanolett.6b05152} {\bibfield  {journal}
  {\bibinfo  {journal} {Nano Letters}\ }\textbf {\bibinfo {volume} {17}},\
  \bibinfo {pages} {2825} (\bibinfo {year} {2017})}\BibitemShut {NoStop}%
\bibitem [{\citenamefont {Heron}\ \emph {et~al.}(2011)\citenamefont {Heron},
  \citenamefont {Trassin}, \citenamefont {Ashraf}, \citenamefont {Gajek},
  \citenamefont {He}, \citenamefont {Yang}, \citenamefont {Nikonov},
  \citenamefont {Chu}, \citenamefont {Salahuddin},\ and\ \citenamefont
  {Ramesh}}]{PhysRevLett.107.217202}%
  \BibitemOpen
  \bibfield  {author} {\bibinfo {author} {\bibfnamefont {J.~T.}\ \bibnamefont
  {Heron}}, \bibinfo {author} {\bibfnamefont {M.}~\bibnamefont {Trassin}},
  \bibinfo {author} {\bibfnamefont {K.}~\bibnamefont {Ashraf}}, \bibinfo
  {author} {\bibfnamefont {M.}~\bibnamefont {Gajek}}, \bibinfo {author}
  {\bibfnamefont {Q.}~\bibnamefont {He}}, \bibinfo {author} {\bibfnamefont
  {S.~Y.}\ \bibnamefont {Yang}}, \bibinfo {author} {\bibfnamefont {D.~E.}\
  \bibnamefont {Nikonov}}, \bibinfo {author} {\bibfnamefont {Y.-H.}\
  \bibnamefont {Chu}}, \bibinfo {author} {\bibfnamefont {S.}~\bibnamefont
  {Salahuddin}}, \ and\ \bibinfo {author} {\bibfnamefont {R.}~\bibnamefont
  {Ramesh}},\ }\href {\doibase 10.1103/PhysRevLett.107.217202} {\bibfield
  {journal} {\bibinfo  {journal} {Phys. Rev. Lett.}\ }\textbf {\bibinfo
  {volume} {107}},\ \bibinfo {pages} {217202} (\bibinfo {year}
  {2011})}\BibitemShut {NoStop}%
\bibitem [{\citenamefont {Heron}\ \emph
  {et~al.}(2014{\natexlab{b}})\citenamefont {Heron}, \citenamefont {Schlom},\
  and\ \citenamefont {Ramesh}}]{doi:10.1063/1.4870957}%
  \BibitemOpen
  \bibfield  {author} {\bibinfo {author} {\bibfnamefont {J.~T.}\ \bibnamefont
  {Heron}}, \bibinfo {author} {\bibfnamefont {D.~G.}\ \bibnamefont {Schlom}}, \
  and\ \bibinfo {author} {\bibfnamefont {R.}~\bibnamefont {Ramesh}},\ }\href
  {\doibase 10.1063/1.4870957} {\bibfield  {journal} {\bibinfo  {journal}
  {Applied Physics Reviews}\ }\textbf {\bibinfo {volume} {1}},\ \bibinfo
  {pages} {021303} (\bibinfo {year} {2014}{\natexlab{b}})}\BibitemShut
  {NoStop}%
\bibitem [{\citenamefont {Song}\ \emph {et~al.}(2017)\citenamefont {Song},
  \citenamefont {Cui}, \citenamefont {Li}, \citenamefont {Zhou},\ and\
  \citenamefont {Pan}}]{SONG201733_review}%
  \BibitemOpen
  \bibfield  {author} {\bibinfo {author} {\bibfnamefont {C.}~\bibnamefont
  {Song}}, \bibinfo {author} {\bibfnamefont {B.}~\bibnamefont {Cui}}, \bibinfo
  {author} {\bibfnamefont {F.}~\bibnamefont {Li}}, \bibinfo {author}
  {\bibfnamefont {X.}~\bibnamefont {Zhou}}, \ and\ \bibinfo {author}
  {\bibfnamefont {F.}~\bibnamefont {Pan}},\ }\href {\doibase
  https://doi.org/10.1016/j.pmatsci.2017.02.002} {\bibfield  {journal}
  {\bibinfo  {journal} {Progress in Materials Science}\ }\textbf {\bibinfo
  {volume} {87}},\ \bibinfo {pages} {33 } (\bibinfo {year} {2017})}\BibitemShut
  {NoStop}%
\bibitem [{\citenamefont {Spaldin}\ and\ \citenamefont
  {Ramesh}(2019)}]{Spaldin_NR_2019}%
  \BibitemOpen
  \bibfield  {author} {\bibinfo {author} {\bibfnamefont {N.~A.}\ \bibnamefont
  {Spaldin}}\ and\ \bibinfo {author} {\bibfnamefont {R.}~\bibnamefont
  {Ramesh}},\ }\href {\doibase https://doi.org/10.1038/s41563-018-0275-2}
  {\bibfield  {journal} {\bibinfo  {journal} {Nature materials}\ }\textbf
  {\bibinfo {volume} {18}},\ \bibinfo {pages} {203} (\bibinfo {year}
  {2019})}\BibitemShut {NoStop}%
\bibitem [{\citenamefont {Wu}\ \emph {et~al.}(2010)\citenamefont {Wu},
  \citenamefont {Cybart}, \citenamefont {Yu}, \citenamefont {Rossell},
  \citenamefont {Zhang}, \citenamefont {Ramesh},\ and\ \citenamefont
  {Dynes}}]{Wu_NM_2010}%
  \BibitemOpen
  \bibfield  {author} {\bibinfo {author} {\bibfnamefont {S.~M.}\ \bibnamefont
  {Wu}}, \bibinfo {author} {\bibfnamefont {S.~A.}\ \bibnamefont {Cybart}},
  \bibinfo {author} {\bibfnamefont {P.}~\bibnamefont {Yu}}, \bibinfo {author}
  {\bibfnamefont {M.~D.}\ \bibnamefont {Rossell}}, \bibinfo {author}
  {\bibfnamefont {J.~X.}\ \bibnamefont {Zhang}}, \bibinfo {author}
  {\bibfnamefont {R.}~\bibnamefont {Ramesh}}, \ and\ \bibinfo {author}
  {\bibfnamefont {R.~C.}\ \bibnamefont {Dynes}},\ }\href {\doibase
  https://doi.org/10.1038/nmat2803} {\bibfield  {journal} {\bibinfo  {journal}
  {Nature materials}\ }\textbf {\bibinfo {volume} {9}},\ \bibinfo {pages} {756}
  (\bibinfo {year} {2010})}\BibitemShut {NoStop}%
\bibitem [{\citenamefont {He}\ \emph {et~al.}(2010)\citenamefont {He},
  \citenamefont {Wang}, \citenamefont {Wu}, \citenamefont {Caruso},
  \citenamefont {Vescovo}, \citenamefont {Belashchenko}, \citenamefont
  {Dowben},\ and\ \citenamefont {Binek}}]{He_NM_2010}%
  \BibitemOpen
  \bibfield  {author} {\bibinfo {author} {\bibfnamefont {X.}~\bibnamefont
  {He}}, \bibinfo {author} {\bibfnamefont {Y.}~\bibnamefont {Wang}}, \bibinfo
  {author} {\bibfnamefont {N.}~\bibnamefont {Wu}}, \bibinfo {author}
  {\bibfnamefont {A.~N.}\ \bibnamefont {Caruso}}, \bibinfo {author}
  {\bibfnamefont {E.}~\bibnamefont {Vescovo}}, \bibinfo {author} {\bibfnamefont
  {K.~D.}\ \bibnamefont {Belashchenko}}, \bibinfo {author} {\bibfnamefont
  {P.~A.}\ \bibnamefont {Dowben}}, \ and\ \bibinfo {author} {\bibfnamefont
  {C.}~\bibnamefont {Binek}},\ }\href {\doibase
  https://doi.org/10.1038/nmat2785} {\bibfield  {journal} {\bibinfo  {journal}
  {Nature materials}\ }\textbf {\bibinfo {volume} {9}},\ \bibinfo {pages} {579}
  (\bibinfo {year} {2010})}\BibitemShut {NoStop}%
\bibitem [{\citenamefont {Scherwitzl}\ \emph {et~al.}(2010)\citenamefont
  {Scherwitzl}, \citenamefont {Zubko}, \citenamefont {Lezama}, \citenamefont
  {Ono}, \citenamefont {Morpurgo}, \citenamefont {Catalan},\ and\ \citenamefont
  {Triscone}}]{doi:10.1002/adma.201003241}%
  \BibitemOpen
  \bibfield  {author} {\bibinfo {author} {\bibfnamefont {R.}~\bibnamefont
  {Scherwitzl}}, \bibinfo {author} {\bibfnamefont {P.}~\bibnamefont {Zubko}},
  \bibinfo {author} {\bibfnamefont {I.~G.}\ \bibnamefont {Lezama}}, \bibinfo
  {author} {\bibfnamefont {S.}~\bibnamefont {Ono}}, \bibinfo {author}
  {\bibfnamefont {A.~F.}\ \bibnamefont {Morpurgo}}, \bibinfo {author}
  {\bibfnamefont {G.}~\bibnamefont {Catalan}}, \ and\ \bibinfo {author}
  {\bibfnamefont {J.-M.}\ \bibnamefont {Triscone}},\ }\href {\doibase
  10.1002/adma.201003241} {\bibfield  {journal} {\bibinfo  {journal} {Advanced
  Materials}\ }\textbf {\bibinfo {volume} {22}},\ \bibinfo {pages} {5517}
  (\bibinfo {year} {2010})}\BibitemShut {NoStop}%
\bibitem [{\citenamefont {Ueno}\ \emph {et~al.}(2008)\citenamefont {Ueno},
  \citenamefont {Nakamura}, \citenamefont {Shimotani}, \citenamefont {Ohtomo},
  \citenamefont {Kimura}, \citenamefont {Nojima}, \citenamefont {Aoki},
  \citenamefont {Iwasa},\ and\ \citenamefont {Kawasaki}}]{Ueno2008}%
  \BibitemOpen
  \bibfield  {author} {\bibinfo {author} {\bibfnamefont {K.}~\bibnamefont
  {Ueno}}, \bibinfo {author} {\bibfnamefont {S.}~\bibnamefont {Nakamura}},
  \bibinfo {author} {\bibfnamefont {H.}~\bibnamefont {Shimotani}}, \bibinfo
  {author} {\bibfnamefont {A.}~\bibnamefont {Ohtomo}}, \bibinfo {author}
  {\bibfnamefont {N.}~\bibnamefont {Kimura}}, \bibinfo {author} {\bibfnamefont
  {T.}~\bibnamefont {Nojima}}, \bibinfo {author} {\bibfnamefont
  {H.}~\bibnamefont {Aoki}}, \bibinfo {author} {\bibfnamefont {Y.}~\bibnamefont
  {Iwasa}}, \ and\ \bibinfo {author} {\bibfnamefont {M.}~\bibnamefont
  {Kawasaki}},\ }\href {\doibase https://doi.org/10.1038/nmat2298} {\bibfield
  {journal} {\bibinfo  {journal} {Nature materials}\ }\textbf {\bibinfo
  {volume} {7}},\ \bibinfo {pages} {855} (\bibinfo {year} {2008})}\BibitemShut
  {NoStop}%
\bibitem [{\citenamefont {Caviglia}\ \emph {et~al.}(2008)\citenamefont
  {Caviglia}, \citenamefont {Gariglio}, \citenamefont {Reyren}, \citenamefont
  {Jaccard}, \citenamefont {Schneider}, \citenamefont {Gabay}, \citenamefont
  {Thiel}, \citenamefont {Hammerl}, \citenamefont {Mannhart},\ and\
  \citenamefont {Triscone}}]{Caviglia_2008}%
  \BibitemOpen
  \bibfield  {author} {\bibinfo {author} {\bibfnamefont {A.~D.}\ \bibnamefont
  {Caviglia}}, \bibinfo {author} {\bibfnamefont {S.}~\bibnamefont {Gariglio}},
  \bibinfo {author} {\bibfnamefont {N.}~\bibnamefont {Reyren}}, \bibinfo
  {author} {\bibfnamefont {D.}~\bibnamefont {Jaccard}}, \bibinfo {author}
  {\bibfnamefont {T.}~\bibnamefont {Schneider}}, \bibinfo {author}
  {\bibfnamefont {M.}~\bibnamefont {Gabay}}, \bibinfo {author} {\bibfnamefont
  {S.}~\bibnamefont {Thiel}}, \bibinfo {author} {\bibfnamefont
  {G.}~\bibnamefont {Hammerl}}, \bibinfo {author} {\bibfnamefont
  {J.}~\bibnamefont {Mannhart}}, \ and\ \bibinfo {author} {\bibfnamefont
  {J.~M.}\ \bibnamefont {Triscone}},\ }\href {\doibase
  https://doi.org/10.1038/nature07576} {\bibfield  {journal} {\bibinfo
  {journal} {Nature}\ }\textbf {\bibinfo {volume} {456}},\ \bibinfo {pages}
  {624} (\bibinfo {year} {2008})}\BibitemShut {NoStop}%
\bibitem [{\citenamefont {Newaz}\ \emph {et~al.}(2013)\citenamefont {Newaz},
  \citenamefont {Prasai}, \citenamefont {Ziegler}, \citenamefont {Caudel},
  \citenamefont {Robinson}, \citenamefont {Jr.]},\ and\ \citenamefont
  {Bolotin}}]{NEWAZ201349}%
  \BibitemOpen
  \bibfield  {author} {\bibinfo {author} {\bibfnamefont {A.}~\bibnamefont
  {Newaz}}, \bibinfo {author} {\bibfnamefont {D.}~\bibnamefont {Prasai}},
  \bibinfo {author} {\bibfnamefont {J.}~\bibnamefont {Ziegler}}, \bibinfo
  {author} {\bibfnamefont {D.}~\bibnamefont {Caudel}}, \bibinfo {author}
  {\bibfnamefont {S.}~\bibnamefont {Robinson}}, \bibinfo {author}
  {\bibfnamefont {R.~H.}\ \bibnamefont {Jr.]}}, \ and\ \bibinfo {author}
  {\bibfnamefont {K.}~\bibnamefont {Bolotin}},\ }\href {\doibase
  https://doi.org/10.1016/j.ssc.2012.11.010} {\bibfield  {journal} {\bibinfo
  {journal} {Solid State Communications}\ }\textbf {\bibinfo {volume} {155}},\
  \bibinfo {pages} {49 } (\bibinfo {year} {2013})}\BibitemShut {NoStop}%
\bibitem [{\citenamefont {Wang}\ \emph {et~al.}(2014)\citenamefont {Wang},
  \citenamefont {Webber}, \citenamefont {Hudspeth}, \citenamefont
  {Hinterstein},\ and\ \citenamefont {Daniels}}]{doi:10.1063/1.4898573}%
  \BibitemOpen
  \bibfield  {author} {\bibinfo {author} {\bibfnamefont {Z.}~\bibnamefont
  {Wang}}, \bibinfo {author} {\bibfnamefont {K.~G.}\ \bibnamefont {Webber}},
  \bibinfo {author} {\bibfnamefont {J.~M.}\ \bibnamefont {Hudspeth}}, \bibinfo
  {author} {\bibfnamefont {M.}~\bibnamefont {Hinterstein}}, \ and\ \bibinfo
  {author} {\bibfnamefont {J.~E.}\ \bibnamefont {Daniels}},\ }\href {\doibase
  10.1063/1.4898573} {\bibfield  {journal} {\bibinfo  {journal} {Applied
  Physics Letters}\ }\textbf {\bibinfo {volume} {105}},\ \bibinfo {pages}
  {161903} (\bibinfo {year} {2014})}\BibitemShut {NoStop}%
\bibitem [{\citenamefont {Cao}\ \emph {et~al.}(2018)\citenamefont {Cao},
  \citenamefont {Terzic}, \citenamefont {Zhao}, \citenamefont {Zheng},
  \citenamefont {De~Long},\ and\ \citenamefont
  {Riseborough}}]{PhysRevLett.120.017201}%
  \BibitemOpen
  \bibfield  {author} {\bibinfo {author} {\bibfnamefont {G.}~\bibnamefont
  {Cao}}, \bibinfo {author} {\bibfnamefont {J.}~\bibnamefont {Terzic}},
  \bibinfo {author} {\bibfnamefont {H.~D.}\ \bibnamefont {Zhao}}, \bibinfo
  {author} {\bibfnamefont {H.}~\bibnamefont {Zheng}}, \bibinfo {author}
  {\bibfnamefont {L.~E.}\ \bibnamefont {De~Long}}, \ and\ \bibinfo {author}
  {\bibfnamefont {P.~S.}\ \bibnamefont {Riseborough}},\ }\href {\doibase
  10.1103/PhysRevLett.120.017201} {\bibfield  {journal} {\bibinfo  {journal}
  {Phys. Rev. Lett.}\ }\textbf {\bibinfo {volume} {120}},\ \bibinfo {pages}
  {017201} (\bibinfo {year} {2018})}\BibitemShut {NoStop}%
\bibitem [{\citenamefont {Yi}\ \emph {et~al.}(2020)\citenamefont {Yi},
  \citenamefont {Wang}, \citenamefont {van't Erve}, \citenamefont {Xu},
  \citenamefont {Yuan}, \citenamefont {Veit}, \citenamefont {Balakrishnan},
  \citenamefont {Choi}, \citenamefont {N'Diaye}, \citenamefont {Shafer},
  \citenamefont {Arenholz}, \citenamefont {Grutter}, \citenamefont {Xu},
  \citenamefont {Yu}, \citenamefont {Jonker},\ and\ \citenamefont
  {Suzuki}}]{Yi_NC2020}%
  \BibitemOpen
  \bibfield  {author} {\bibinfo {author} {\bibfnamefont {D.}~\bibnamefont
  {Yi}}, \bibinfo {author} {\bibfnamefont {Y.}~\bibnamefont {Wang}}, \bibinfo
  {author} {\bibfnamefont {O.~M.~J.}\ \bibnamefont {van't Erve}}, \bibinfo
  {author} {\bibfnamefont {L.}~\bibnamefont {Xu}}, \bibinfo {author}
  {\bibfnamefont {H.}~\bibnamefont {Yuan}}, \bibinfo {author} {\bibfnamefont
  {M.~J.}\ \bibnamefont {Veit}}, \bibinfo {author} {\bibfnamefont {P.~P.}\
  \bibnamefont {Balakrishnan}}, \bibinfo {author} {\bibfnamefont
  {Y.}~\bibnamefont {Choi}}, \bibinfo {author} {\bibfnamefont {A.~T.}\
  \bibnamefont {N'Diaye}}, \bibinfo {author} {\bibfnamefont
  {P.}~\bibnamefont {Shafer}}, \bibinfo {author} {\bibfnamefont
  {E.}~\bibnamefont {Arenholz}}, \bibinfo {author} {\bibfnamefont
  {A.}~\bibnamefont {Grutter}}, \bibinfo {author} {\bibfnamefont
  {H.}~\bibnamefont {Xu}}, \bibinfo {author} {\bibfnamefont {P.}~\bibnamefont
  {Yu}}, \bibinfo {author} {\bibfnamefont {B.~T.}\ \bibnamefont {Jonker}}, \
  and\ \bibinfo {author} {\bibfnamefont {Y.}~\bibnamefont {Suzuki}},\ }\href
  {\doibase https://doi.org/10.1038/s41467-020-14631-3} {\bibfield  {journal}
  {\bibinfo  {journal} {Nat Commun}\ }\textbf {\bibinfo {volume} {11}},\
  \bibinfo {pages} {902} (\bibinfo {year} {2020})}\BibitemShut {NoStop}%
\bibitem [{\citenamefont {Nenciu}(1991)}]{RevModPhys.63.91}%
  \BibitemOpen
  \bibfield  {author} {\bibinfo {author} {\bibfnamefont {G.}~\bibnamefont
  {Nenciu}},\ }\href {\doibase 10.1103/RevModPhys.63.91} {\bibfield  {journal}
  {\bibinfo  {journal} {Rev. Mod. Phys.}\ }\textbf {\bibinfo {volume} {63}},\
  \bibinfo {pages} {91} (\bibinfo {year} {1991})}\BibitemShut {NoStop}%
\bibitem [{\citenamefont {Kane}(1960)}]{KANE1960181}%
  \BibitemOpen
  \bibfield  {author} {\bibinfo {author} {\bibfnamefont {E.}~\bibnamefont
  {Kane}},\ }\href {\doibase https://doi.org/10.1016/0022-3697(60)90035-4}
  {\bibfield  {journal} {\bibinfo  {journal} {Journal of Physics and Chemistry
  of Solids}\ }\textbf {\bibinfo {volume} {12}},\ \bibinfo {pages} {181 }
  (\bibinfo {year} {1960})}\BibitemShut {NoStop}%
\bibitem [{\citenamefont {Souza}\ \emph {et~al.}(2002)\citenamefont {Souza},
  \citenamefont {\'I\~niguez},\ and\ \citenamefont
  {Vanderbilt}}]{PhysRevLett.89.117602}%
  \BibitemOpen
  \bibfield  {author} {\bibinfo {author} {\bibfnamefont {I.}~\bibnamefont
  {Souza}}, \bibinfo {author} {\bibfnamefont {J.}~\bibnamefont {\'I\~niguez}},
  \ and\ \bibinfo {author} {\bibfnamefont {D.}~\bibnamefont {Vanderbilt}},\
  }\href {\doibase 10.1103/PhysRevLett.89.117602} {\bibfield  {journal}
  {\bibinfo  {journal} {Phys. Rev. Lett.}\ }\textbf {\bibinfo {volume} {89}},\
  \bibinfo {pages} {117602} (\bibinfo {year} {2002})}\BibitemShut {NoStop}%
\bibitem [{\citenamefont {Nunes}\ and\ \citenamefont
  {Gonze}(2001)}]{PhysRevB.63.155107}%
  \BibitemOpen
  \bibfield  {author} {\bibinfo {author} {\bibfnamefont {R.~W.}\ \bibnamefont
  {Nunes}}\ and\ \bibinfo {author} {\bibfnamefont {X.}~\bibnamefont {Gonze}},\
  }\href {\doibase 10.1103/PhysRevB.63.155107} {\bibfield  {journal} {\bibinfo
  {journal} {Phys. Rev. B}\ }\textbf {\bibinfo {volume} {63}},\ \bibinfo
  {pages} {155107} (\bibinfo {year} {2001})}\BibitemShut {NoStop}%
\bibitem [{\citenamefont {Umari}\ and\ \citenamefont
  {Pasquarello}(2002)}]{PhysRevLett.89.157602}%
  \BibitemOpen
  \bibfield  {author} {\bibinfo {author} {\bibfnamefont {P.}~\bibnamefont
  {Umari}}\ and\ \bibinfo {author} {\bibfnamefont {A.}~\bibnamefont
  {Pasquarello}},\ }\href {\doibase 10.1103/PhysRevLett.89.157602} {\bibfield
  {journal} {\bibinfo  {journal} {Phys. Rev. Lett.}\ }\textbf {\bibinfo
  {volume} {89}},\ \bibinfo {pages} {157602} (\bibinfo {year}
  {2002})}\BibitemShut {NoStop}%
\bibitem [{\citenamefont {Sai}\ \emph {et~al.}(2002)\citenamefont {Sai},
  \citenamefont {Rabe},\ and\ \citenamefont {Vanderbilt}}]{PhysRevB.66.104108}%
  \BibitemOpen
  \bibfield  {author} {\bibinfo {author} {\bibfnamefont {N.}~\bibnamefont
  {Sai}}, \bibinfo {author} {\bibfnamefont {K.~M.}\ \bibnamefont {Rabe}}, \
  and\ \bibinfo {author} {\bibfnamefont {D.}~\bibnamefont {Vanderbilt}},\
  }\href {\doibase 10.1103/PhysRevB.66.104108} {\bibfield  {journal} {\bibinfo
  {journal} {Phys. Rev. B}\ }\textbf {\bibinfo {volume} {66}},\ \bibinfo
  {pages} {104108} (\bibinfo {year} {2002})}\BibitemShut {NoStop}%
\bibitem [{\citenamefont {King-Smith}\ and\ \citenamefont
  {Vanderbilt}(1993)}]{PhysRevB.47.1651}%
  \BibitemOpen
  \bibfield  {author} {\bibinfo {author} {\bibfnamefont {R.~D.}\ \bibnamefont
  {King-Smith}}\ and\ \bibinfo {author} {\bibfnamefont {D.}~\bibnamefont
  {Vanderbilt}},\ }\href {\doibase 10.1103/PhysRevB.47.1651} {\bibfield
  {journal} {\bibinfo  {journal} {Phys. Rev. B}\ }\textbf {\bibinfo {volume}
  {47}},\ \bibinfo {pages} {1651} (\bibinfo {year} {1993})}\BibitemShut
  {NoStop}%
\bibitem [{\citenamefont {Resta}(1994)}]{RevModPhys.66.899}%
  \BibitemOpen
  \bibfield  {author} {\bibinfo {author} {\bibfnamefont {R.}~\bibnamefont
  {Resta}},\ }\href {\doibase 10.1103/RevModPhys.66.899} {\bibfield  {journal}
  {\bibinfo  {journal} {Rev. Mod. Phys.}\ }\textbf {\bibinfo {volume} {66}},\
  \bibinfo {pages} {899} (\bibinfo {year} {1994})}\BibitemShut {NoStop}%
\bibitem [{\citenamefont {Vanderbilt}\ and\ \citenamefont
  {King-Smith}(1998)}]{David1998_USPP}%
  \BibitemOpen
  \bibfield  {author} {\bibinfo {author} {\bibfnamefont {D.}~\bibnamefont
  {Vanderbilt}}\ and\ \bibinfo {author} {\bibfnamefont {R.}~\bibnamefont
  {King-Smith}},\ }\href {https://arxiv.org/abs/cond-mat/9801177v1} {\bibfield
  {journal} {\bibinfo  {journal} {arXiv:cond-mat/9801177}\ } (\bibinfo {year}
  {1998})}\BibitemShut {NoStop}%
\bibitem [{\citenamefont {Zwanziger}\ \emph {et~al.}(2012)\citenamefont
  {Zwanziger}, \citenamefont {Galbraith}, \citenamefont {Kipouros},
  \citenamefont {Torrent}, \citenamefont {Giantomassi},\ and\ \citenamefont
  {Gonze}}]{ZWANZIGER2012113}%
  \BibitemOpen
  \bibfield  {author} {\bibinfo {author} {\bibfnamefont {J.}~\bibnamefont
  {Zwanziger}}, \bibinfo {author} {\bibfnamefont {J.}~\bibnamefont
  {Galbraith}}, \bibinfo {author} {\bibfnamefont {Y.}~\bibnamefont {Kipouros}},
  \bibinfo {author} {\bibfnamefont {M.}~\bibnamefont {Torrent}}, \bibinfo
  {author} {\bibfnamefont {M.}~\bibnamefont {Giantomassi}}, \ and\ \bibinfo
  {author} {\bibfnamefont {X.}~\bibnamefont {Gonze}},\ }\href {\doibase
  https://doi.org/10.1016/j.commatsci.2012.01.028} {\bibfield  {journal}
  {\bibinfo  {journal} {Computational Materials Science}\ }\textbf {\bibinfo
  {volume} {58}},\ \bibinfo {pages} {113 } (\bibinfo {year}
  {2012})}\BibitemShut {NoStop}%
\bibitem [{\citenamefont {Veithen}\ \emph {et~al.}(2005)\citenamefont
  {Veithen}, \citenamefont {Gonze},\ and\ \citenamefont
  {Ghosez}}]{PhysRevB.71.125107}%
  \BibitemOpen
  \bibfield  {author} {\bibinfo {author} {\bibfnamefont {M.}~\bibnamefont
  {Veithen}}, \bibinfo {author} {\bibfnamefont {X.}~\bibnamefont {Gonze}}, \
  and\ \bibinfo {author} {\bibfnamefont {P.}~\bibnamefont {Ghosez}},\ }\href
  {\doibase 10.1103/PhysRevB.71.125107} {\bibfield  {journal} {\bibinfo
  {journal} {Phys. Rev. B}\ }\textbf {\bibinfo {volume} {71}},\ \bibinfo
  {pages} {125107} (\bibinfo {year} {2005})}\BibitemShut {NoStop}%
\bibitem [{\citenamefont {Roman}\ \emph {et~al.}(2006)\citenamefont {Roman},
  \citenamefont {Yates}, \citenamefont {Veithen}, \citenamefont {Vanderbilt},\
  and\ \citenamefont {Souza}}]{PhysRevB.74.245204}%
  \BibitemOpen
  \bibfield  {author} {\bibinfo {author} {\bibfnamefont {E.}~\bibnamefont
  {Roman}}, \bibinfo {author} {\bibfnamefont {J.~R.}\ \bibnamefont {Yates}},
  \bibinfo {author} {\bibfnamefont {M.}~\bibnamefont {Veithen}}, \bibinfo
  {author} {\bibfnamefont {D.}~\bibnamefont {Vanderbilt}}, \ and\ \bibinfo
  {author} {\bibfnamefont {I.}~\bibnamefont {Souza}},\ }\href {\doibase
  10.1103/PhysRevB.74.245204} {\bibfield  {journal} {\bibinfo  {journal} {Phys.
  Rev. B}\ }\textbf {\bibinfo {volume} {74}},\ \bibinfo {pages} {245204}
  (\bibinfo {year} {2006})}\BibitemShut {NoStop}%
\bibitem [{\citenamefont {de~Gironcoli}\ \emph {et~al.}(1989)\citenamefont
  {de~Gironcoli}, \citenamefont {Baroni},\ and\ \citenamefont
  {Resta}}]{PhysRevLett.62.2853}%
  \BibitemOpen
  \bibfield  {author} {\bibinfo {author} {\bibfnamefont {S.}~\bibnamefont
  {de~Gironcoli}}, \bibinfo {author} {\bibfnamefont {S.}~\bibnamefont
  {Baroni}}, \ and\ \bibinfo {author} {\bibfnamefont {R.}~\bibnamefont
  {Resta}},\ }\href {\doibase 10.1103/PhysRevLett.62.2853} {\bibfield
  {journal} {\bibinfo  {journal} {Phys. Rev. Lett.}\ }\textbf {\bibinfo
  {volume} {62}},\ \bibinfo {pages} {2853} (\bibinfo {year}
  {1989})}\BibitemShut {NoStop}%
\bibitem [{\citenamefont {Minami}\ \emph {et~al.}(2024)\citenamefont {Minami},
  \citenamefont {Nakayama},\ and\ \citenamefont
  {Shimada}}]{MinamiNakayamaShimada2024}%
  \BibitemOpen
  \bibfield  {author} {\bibinfo {author} {\bibfnamefont {S.}~\bibnamefont
  {Minami}}, \bibinfo {author} {\bibfnamefont {T.}~\bibnamefont {Nakayama}}, \
  and\ \bibinfo {author} {\bibfnamefont {T.}~\bibnamefont {Shimada}},\ }\href
  {\doibase 10.1016/j.actamat.2024.120248} {\bibfield  {journal} {\bibinfo
  {journal} {Acta Materialia}\ }\textbf {\bibinfo {volume} {278}},\ \bibinfo
  {pages} {120248} (\bibinfo {year} {2024})}\BibitemShut {NoStop}%
\bibitem [{\citenamefont {Wang}\ and\ \citenamefont
  {Vanderbilt}(2007)}]{PhysRevB.75.115116}%
  \BibitemOpen
  \bibfield  {author} {\bibinfo {author} {\bibfnamefont {X.}~\bibnamefont
  {Wang}}\ and\ \bibinfo {author} {\bibfnamefont {D.}~\bibnamefont
  {Vanderbilt}},\ }\href {\doibase 10.1103/PhysRevB.75.115116} {\bibfield
  {journal} {\bibinfo  {journal} {Phys. Rev. B}\ }\textbf {\bibinfo {volume}
  {75}},\ \bibinfo {pages} {115116} (\bibinfo {year} {2007})}\BibitemShut
  {NoStop}%
\bibitem [{\citenamefont {Zhao}\ \emph {et~al.}(2025)\citenamefont {Zhao},
  \citenamefont {Fu}, \citenamefont {Yang}, \citenamefont {Wang}, \citenamefont
  {Bellaiche},\ and\ \citenamefont {Ma}}]{PhysRevLett.134.046801}%
  \BibitemOpen
  \bibfield  {author} {\bibinfo {author} {\bibfnamefont {H.~J.}\ \bibnamefont
  {Zhao}}, \bibinfo {author} {\bibfnamefont {Y.}~\bibnamefont {Fu}}, \bibinfo
  {author} {\bibfnamefont {Y.}~\bibnamefont {Yang}}, \bibinfo {author}
  {\bibfnamefont {Y.}~\bibnamefont {Wang}}, \bibinfo {author} {\bibfnamefont
  {L.}~\bibnamefont {Bellaiche}}, \ and\ \bibinfo {author} {\bibfnamefont
  {Y.}~\bibnamefont {Ma}},\ }\href {\doibase 10.1103/PhysRevLett.134.046801}
  {\bibfield  {journal} {\bibinfo  {journal} {Phys. Rev. Lett.}\ }\textbf
  {\bibinfo {volume} {134}},\ \bibinfo {pages} {046801} (\bibinfo {year}
  {2025})}\BibitemShut {NoStop}%
\bibitem [{\citenamefont {Umari}\ and\ \citenamefont
  {Pasquarello}(2007)}]{PhysRevLett.98.176402}%
  \BibitemOpen
  \bibfield  {author} {\bibinfo {author} {\bibfnamefont {P.}~\bibnamefont
  {Umari}}\ and\ \bibinfo {author} {\bibfnamefont {A.}~\bibnamefont
  {Pasquarello}},\ }\href {\doibase 10.1103/PhysRevLett.98.176402} {\bibfield
  {journal} {\bibinfo  {journal} {Phys. Rev. Lett.}\ }\textbf {\bibinfo
  {volume} {98}},\ \bibinfo {pages} {176402} (\bibinfo {year}
  {2007})}\BibitemShut {NoStop}%
\bibitem [{\citenamefont {Yu}\ \emph {et~al.}(2025)\citenamefont {Yu},
  \citenamefont {Zhao}, \citenamefont {Bellaiche},\ and\ \citenamefont
  {Ma}}]{96gy-sn83}%
  \BibitemOpen
  \bibfield  {author} {\bibinfo {author} {\bibfnamefont {L.}~\bibnamefont
  {Yu}}, \bibinfo {author} {\bibfnamefont {H.~J.}\ \bibnamefont {Zhao}},
  \bibinfo {author} {\bibfnamefont {L.}~\bibnamefont {Bellaiche}}, \ and\
  \bibinfo {author} {\bibfnamefont {Y.}~\bibnamefont {Ma}},\ }\href {\doibase
  10.1103/96gy-sn83} {\bibfield  {journal} {\bibinfo  {journal} {Phys. Rev.
  Lett.}\ }\textbf {\bibinfo {volume} {135}},\ \bibinfo {pages} {256704}
  (\bibinfo {year} {2025})}\BibitemShut {NoStop}%
\bibitem [{\citenamefont {Moriwake}\ \emph {et~al.}(2016)\citenamefont
  {Moriwake}, \citenamefont {Konishi}, \citenamefont {Ogawa}, \citenamefont
  {Fisher}, \citenamefont {Kuwabara},\ and\ \citenamefont
  {Fu}}]{doi:10.1063/1.4941319}%
  \BibitemOpen
  \bibfield  {author} {\bibinfo {author} {\bibfnamefont {H.}~\bibnamefont
  {Moriwake}}, \bibinfo {author} {\bibfnamefont {A.}~\bibnamefont {Konishi}},
  \bibinfo {author} {\bibfnamefont {T.}~\bibnamefont {Ogawa}}, \bibinfo
  {author} {\bibfnamefont {C.~A.~J.}\ \bibnamefont {Fisher}}, \bibinfo {author}
  {\bibfnamefont {A.}~\bibnamefont {Kuwabara}}, \ and\ \bibinfo {author}
  {\bibfnamefont {D.}~\bibnamefont {Fu}},\ }\href {\doibase 10.1063/1.4941319}
  {\bibfield  {journal} {\bibinfo  {journal} {Journal of Applied Physics}\
  }\textbf {\bibinfo {volume} {119}},\ \bibinfo {pages} {064102} (\bibinfo
  {year} {2016})}\BibitemShut {NoStop}%
\bibitem [{\citenamefont {Fu}\ and\ \citenamefont
  {Bellaiche}(2003)}]{PhysRevLett.91.057601}%
  \BibitemOpen
  \bibfield  {author} {\bibinfo {author} {\bibfnamefont {H.}~\bibnamefont
  {Fu}}\ and\ \bibinfo {author} {\bibfnamefont {L.}~\bibnamefont {Bellaiche}},\
  }\href {\doibase 10.1103/PhysRevLett.91.057601} {\bibfield  {journal}
  {\bibinfo  {journal} {Phys. Rev. Lett.}\ }\textbf {\bibinfo {volume} {91}},\
  \bibinfo {pages} {057601} (\bibinfo {year} {2003})}\BibitemShut {NoStop}%
\bibitem [{\citenamefont {Bilc}\ \emph {et~al.}(2016)\citenamefont {Bilc},
  \citenamefont {Zarbo}, \citenamefont {Garabagiu}, \citenamefont {Bousquet},\
  and\ \citenamefont {Mitoseriu}}]{bilc2016highfieldpropertiestypical}%
  \BibitemOpen
  \bibfield  {author} {\bibinfo {author} {\bibfnamefont {D.~I.}\ \bibnamefont
  {Bilc}}, \bibinfo {author} {\bibfnamefont {L.}~\bibnamefont {Zarbo}},
  \bibinfo {author} {\bibfnamefont {S.}~\bibnamefont {Garabagiu}}, \bibinfo
  {author} {\bibfnamefont {E.}~\bibnamefont {Bousquet}}, \ and\ \bibinfo
  {author} {\bibfnamefont {L.}~\bibnamefont {Mitoseriu}},\ }\href
  {https://arxiv.org/abs/1605.07404} {\enquote {\bibinfo {title} {High field
  properties of typical perovskite ferroelectrics by first-principles
  modeling},}\ } (\bibinfo {year} {2016}),\ \Eprint
  {http://arxiv.org/abs/1605.07404} {arXiv:1605.07404 [cond-mat.mtrl-sci]}
  \BibitemShut {NoStop}%
\bibitem [{\citenamefont {Maeda}\ \emph {et~al.}(2017)\citenamefont {Maeda},
  \citenamefont {Magyari-Kope},\ and\ \citenamefont {Nishi}}]{7939087}%
  \BibitemOpen
  \bibfield  {author} {\bibinfo {author} {\bibfnamefont {T.}~\bibnamefont
  {Maeda}}, \bibinfo {author} {\bibfnamefont {B.}~\bibnamefont {Magyari-Kope}},
  \ and\ \bibinfo {author} {\bibfnamefont {Y.}~\bibnamefont {Nishi}},\ }in\
  \href {\doibase 10.1109/IMW.2017.7939087} {\emph {\bibinfo {booktitle} {2017
  IEEE International Memory Workshop (IMW)}}}\ (\bibinfo {year} {2017})\ pp.\
  \bibinfo {pages} {1--4}\BibitemShut {NoStop}%
\bibitem [{\citenamefont {Ghosh}\ \emph {et~al.}(2019)\citenamefont {Ghosh},
  \citenamefont {Stoji\'c},\ and\ \citenamefont {Binggeli}}]{GHOSH2019166}%
  \BibitemOpen
  \bibfield  {author} {\bibinfo {author} {\bibfnamefont {S.}~\bibnamefont
  {Ghosh}}, \bibinfo {author} {\bibfnamefont {N.}~\bibnamefont {Stoji\'c}}, \
  and\ \bibinfo {author} {\bibfnamefont {N.}~\bibnamefont {Binggeli}},\ }\href
  {\doibase https://doi.org/10.1016/j.physb.2019.06.040} {\bibfield  {journal}
  {\bibinfo  {journal} {Physica B: Condensed Matter}\ }\textbf {\bibinfo
  {volume} {570}},\ \bibinfo {pages} {166} (\bibinfo {year}
  {2019})}\BibitemShut {NoStop}%
\bibitem [{\citenamefont {Chen}\ \emph {et~al.}(2025)\citenamefont {Chen},
  \citenamefont {Xu}, \citenamefont {Jie}, \citenamefont {Feng}, \citenamefont
  {Zhou}, \citenamefont {Wang},\ and\ \citenamefont {Li}}]{Chen2025}%
  \BibitemOpen
  \bibfield  {author} {\bibinfo {author} {\bibfnamefont {J.}~\bibnamefont
  {Chen}}, \bibinfo {author} {\bibfnamefont {M.}~\bibnamefont {Xu}}, \bibinfo
  {author} {\bibfnamefont {Y.}~\bibnamefont {Jie}}, \bibinfo {author}
  {\bibfnamefont {J.}~\bibnamefont {Feng}}, \bibinfo {author} {\bibfnamefont
  {X.}~\bibnamefont {Zhou}}, \bibinfo {author} {\bibfnamefont {Y.}~\bibnamefont
  {Wang}}, \ and\ \bibinfo {author} {\bibfnamefont {Y.}~\bibnamefont {Li}},\
  }\href {\doibase 10.1038/s41467-025-67516-8} {\bibfield  {journal} {\bibinfo
  {journal} {Nature Communications}\ }\textbf {\bibinfo {volume} {17}},\
  \bibinfo {pages} {811} (\bibinfo {year} {2025})}\BibitemShut {NoStop}%
\bibitem [{\citenamefont {Varignon}\ \emph {et~al.}(2016)\citenamefont
  {Varignon}, \citenamefont {Bristowe},\ and\ \citenamefont
  {Ghosez}}]{PhysRevLett.116.057602}%
  \BibitemOpen
  \bibfield  {author} {\bibinfo {author} {\bibfnamefont {J.}~\bibnamefont
  {Varignon}}, \bibinfo {author} {\bibfnamefont {N.~C.}\ \bibnamefont
  {Bristowe}}, \ and\ \bibinfo {author} {\bibfnamefont {P.}~\bibnamefont
  {Ghosez}},\ }\href {\doibase 10.1103/PhysRevLett.116.057602} {\bibfield
  {journal} {\bibinfo  {journal} {Phys. Rev. Lett.}\ }\textbf {\bibinfo
  {volume} {116}},\ \bibinfo {pages} {057602} (\bibinfo {year}
  {2016})}\BibitemShut {NoStop}%
\bibitem [{\citenamefont {Chen}\ \emph {et~al.}(2019)\citenamefont {Chen},
  \citenamefont {Xu}, \citenamefont {Tian}, \citenamefont {Xiang},
  \citenamefont {\'I\~niguez}, \citenamefont {Yang},\ and\ \citenamefont
  {Bellaiche}}]{PhysRevLett.122.247701}%
  \BibitemOpen
  \bibfield  {author} {\bibinfo {author} {\bibfnamefont {L.}~\bibnamefont
  {Chen}}, \bibinfo {author} {\bibfnamefont {C.}~\bibnamefont {Xu}}, \bibinfo
  {author} {\bibfnamefont {H.}~\bibnamefont {Tian}}, \bibinfo {author}
  {\bibfnamefont {H.}~\bibnamefont {Xiang}}, \bibinfo {author} {\bibfnamefont
  {J.}~\bibnamefont {\'I\~niguez}}, \bibinfo {author} {\bibfnamefont
  {Y.}~\bibnamefont {Yang}}, \ and\ \bibinfo {author} {\bibfnamefont
  {L.}~\bibnamefont {Bellaiche}},\ }\href {\doibase
  10.1103/PhysRevLett.122.247701} {\bibfield  {journal} {\bibinfo  {journal}
  {Phys. Rev. Lett.}\ }\textbf {\bibinfo {volume} {122}},\ \bibinfo {pages}
  {247701} (\bibinfo {year} {2019})}\BibitemShut {NoStop}%
\bibitem [{\citenamefont {Zhang}\ \emph {et~al.}(2022)\citenamefont {Zhang},
  \citenamefont {Shen}, \citenamefont {Zhao}, \citenamefont {Dai},
  \citenamefont {Yao}, \citenamefont {Wu}, \citenamefont {Ma}, \citenamefont
  {Nan}, \citenamefont {Wang}, \citenamefont {Yuan}, \citenamefont {da~Silva},
  \citenamefont {Fulanovic}, \citenamefont {Schoekel}, \citenamefont {Liu},
  \citenamefont {Zhang}, \citenamefont {Li}, \citenamefont {Zhang},
  \citenamefont {Wang}, \citenamefont {Roedel},\ and\ \citenamefont
  {Hinterstein}}]{ZhangEtAl2022KNN}%
  \BibitemOpen
  \bibfield  {author} {\bibinfo {author} {\bibfnamefont {M.-H.}\ \bibnamefont
  {Zhang}}, \bibinfo {author} {\bibfnamefont {C.}~\bibnamefont {Shen}},
  \bibinfo {author} {\bibfnamefont {C.}~\bibnamefont {Zhao}}, \bibinfo {author}
  {\bibfnamefont {M.}~\bibnamefont {Dai}}, \bibinfo {author} {\bibfnamefont
  {F.-Z.}\ \bibnamefont {Yao}}, \bibinfo {author} {\bibfnamefont
  {B.}~\bibnamefont {Wu}}, \bibinfo {author} {\bibfnamefont {J.}~\bibnamefont
  {Ma}}, \bibinfo {author} {\bibfnamefont {H.}~\bibnamefont {Nan}}, \bibinfo
  {author} {\bibfnamefont {D.}~\bibnamefont {Wang}}, \bibinfo {author}
  {\bibfnamefont {Q.}~\bibnamefont {Yuan}}, \bibinfo {author} {\bibfnamefont
  {L.~L.}\ \bibnamefont {da~Silva}}, \bibinfo {author} {\bibfnamefont
  {L.}~\bibnamefont {Fulanovic}}, \bibinfo {author} {\bibfnamefont
  {A.}~\bibnamefont {Schoekel}}, \bibinfo {author} {\bibfnamefont
  {P.}~\bibnamefont {Liu}}, \bibinfo {author} {\bibfnamefont {H.}~\bibnamefont
  {Zhang}}, \bibinfo {author} {\bibfnamefont {J.-F.}\ \bibnamefont {Li}},
  \bibinfo {author} {\bibfnamefont {N.}~\bibnamefont {Zhang}}, \bibinfo
  {author} {\bibfnamefont {K.}~\bibnamefont {Wang}}, \bibinfo {author}
  {\bibfnamefont {J.}~\bibnamefont {Roedel}}, \ and\ \bibinfo {author}
  {\bibfnamefont {M.}~\bibnamefont {Hinterstein}},\ }\href {\doibase
  10.1038/s41467-022-31158-x} {\bibfield  {journal} {\bibinfo  {journal}
  {Nature Communications}\ }\textbf {\bibinfo {volume} {13}},\ \bibinfo {pages}
  {3434} (\bibinfo {year} {2022})}\BibitemShut {NoStop}%
\bibitem [{\citenamefont {Arras}\ \emph {et~al.}(2023)\citenamefont {Arras},
  \citenamefont {Paillard},\ and\ \citenamefont
  {Bellaiche}}]{ArrasPaillardBellaiche2023}%
  \BibitemOpen
  \bibfield  {author} {\bibinfo {author} {\bibfnamefont {R.}~\bibnamefont
  {Arras}}, \bibinfo {author} {\bibfnamefont {C.}~\bibnamefont {Paillard}}, \
  and\ \bibinfo {author} {\bibfnamefont {L.}~\bibnamefont {Bellaiche}},\ }\href
  {\doibase 10.1103/PhysRevB.107.144107} {\bibfield  {journal} {\bibinfo
  {journal} {Physical Review B}\ }\textbf {\bibinfo {volume} {107}},\ \bibinfo
  {pages} {144107} (\bibinfo {year} {2023})}\BibitemShut {NoStop}%
\bibitem [{\citenamefont {Han}\ \emph {et~al.}(2024)\citenamefont {Han},
  \citenamefont {Bie}, \citenamefont {Fa}, \citenamefont {Chen}, \citenamefont
  {Tang}, \citenamefont {Guo}, \citenamefont {Xu}, \citenamefont {Ma},
  \citenamefont {Liu}, \citenamefont {Liu}, \citenamefont {Sun},\ and\
  \citenamefont {Luo}}]{HanEtAl2024AFEFE}%
  \BibitemOpen
  \bibfield  {author} {\bibinfo {author} {\bibfnamefont {S.}~\bibnamefont
  {Han}}, \bibinfo {author} {\bibfnamefont {J.}~\bibnamefont {Bie}}, \bibinfo
  {author} {\bibfnamefont {W.}~\bibnamefont {Fa}}, \bibinfo {author}
  {\bibfnamefont {S.}~\bibnamefont {Chen}}, \bibinfo {author} {\bibfnamefont
  {L.}~\bibnamefont {Tang}}, \bibinfo {author} {\bibfnamefont {W.}~\bibnamefont
  {Guo}}, \bibinfo {author} {\bibfnamefont {H.}~\bibnamefont {Xu}}, \bibinfo
  {author} {\bibfnamefont {Y.}~\bibnamefont {Ma}}, \bibinfo {author}
  {\bibfnamefont {Y.}~\bibnamefont {Liu}}, \bibinfo {author} {\bibfnamefont
  {X.}~\bibnamefont {Liu}}, \bibinfo {author} {\bibfnamefont {Z.}~\bibnamefont
  {Sun}}, \ and\ \bibinfo {author} {\bibfnamefont {J.}~\bibnamefont {Luo}},\
  }\href {\doibase 10.1021/jacs.3c13422} {\bibfield  {journal} {\bibinfo
  {journal} {Journal of the American Chemical Society}\ }\textbf {\bibinfo
  {volume} {146}},\ \bibinfo {pages} {8298} (\bibinfo {year}
  {2024})}\BibitemShut {NoStop}%
\bibitem [{\citenamefont {Ghosez}\ and\ \citenamefont
  {Junquera}(2022)}]{GhosezJunquera2022}%
  \BibitemOpen
  \bibfield  {author} {\bibinfo {author} {\bibfnamefont {P.}~\bibnamefont
  {Ghosez}}\ and\ \bibinfo {author} {\bibfnamefont {J.}~\bibnamefont
  {Junquera}},\ }\href {\doibase 10.1146/annurev-conmatphys-040220-045528}
  {\bibfield  {journal} {\bibinfo  {journal} {Annual Review of Condensed Matter
  Physics}\ }\textbf {\bibinfo {volume} {13}},\ \bibinfo {pages} {325}
  (\bibinfo {year} {2022})}\BibitemShut {NoStop}%
\bibitem [{\citenamefont {Gonze}\ \emph {et~al.}(2002)\citenamefont {Gonze},
  \citenamefont {Beuken}, \citenamefont {Caracas}, \citenamefont {Detraux},
  \citenamefont {Fuchs}, \citenamefont {Rignanese}, \citenamefont {Sindic},
  \citenamefont {Verstraete}, \citenamefont {Zerah}, \citenamefont {Jollet},
  \citenamefont {Torrent}, \citenamefont {Roy}, \citenamefont {Mikami},
  \citenamefont {Ghosez}, \citenamefont {Raty},\ and\ \citenamefont
  {Allan}}]{GONZE2002478}%
  \BibitemOpen
  \bibfield  {author} {\bibinfo {author} {\bibfnamefont {X.}~\bibnamefont
  {Gonze}}, \bibinfo {author} {\bibfnamefont {J.-M.}\ \bibnamefont {Beuken}},
  \bibinfo {author} {\bibfnamefont {R.}~\bibnamefont {Caracas}}, \bibinfo
  {author} {\bibfnamefont {F.}~\bibnamefont {Detraux}}, \bibinfo {author}
  {\bibfnamefont {M.}~\bibnamefont {Fuchs}}, \bibinfo {author} {\bibfnamefont
  {G.-M.}\ \bibnamefont {Rignanese}}, \bibinfo {author} {\bibfnamefont
  {L.}~\bibnamefont {Sindic}}, \bibinfo {author} {\bibfnamefont
  {M.}~\bibnamefont {Verstraete}}, \bibinfo {author} {\bibfnamefont
  {G.}~\bibnamefont {Zerah}}, \bibinfo {author} {\bibfnamefont
  {F.}~\bibnamefont {Jollet}}, \bibinfo {author} {\bibfnamefont
  {M.}~\bibnamefont {Torrent}}, \bibinfo {author} {\bibfnamefont
  {A.}~\bibnamefont {Roy}}, \bibinfo {author} {\bibfnamefont {M.}~\bibnamefont
  {Mikami}}, \bibinfo {author} {\bibfnamefont {P.}~\bibnamefont {Ghosez}},
  \bibinfo {author} {\bibfnamefont {J.-Y.}\ \bibnamefont {Raty}}, \ and\
  \bibinfo {author} {\bibfnamefont {D.}~\bibnamefont {Allan}},\ }\href
  {\doibase https://doi.org/10.1016/S0927-0256(02)00325-7} {\bibfield
  {journal} {\bibinfo  {journal} {Computational Materials Science}\ }\textbf
  {\bibinfo {volume} {25}},\ \bibinfo {pages} {478 } (\bibinfo {year}
  {2002})}\BibitemShut {NoStop}%
\bibitem [{\citenamefont {Giannozzi}\ \emph {et~al.}(2009)\citenamefont
  {Giannozzi}, \citenamefont {Baroni}, \citenamefont {Bonini}, \citenamefont
  {Calandra}, \citenamefont {Car}, \citenamefont {Cavazzoni}, \citenamefont
  {Ceresoli}, \citenamefont {Chiarotti}, \citenamefont {Cococcioni},
  \citenamefont {Dabo}, \citenamefont {{Dal Corso}}, \citenamefont
  {de~Gironcoli}, \citenamefont {Fabris}, \citenamefont {Fratesi},
  \citenamefont {Gebauer}, \citenamefont {Gerstmann}, \citenamefont
  {Gougoussis}, \citenamefont {Kokalj}, \citenamefont {Lazzeri}, \citenamefont
  {Martin-Samos}, \citenamefont {Marzari}, \citenamefont {Mauri}, \citenamefont
  {Mazzarello}, \citenamefont {Paolini}, \citenamefont {Pasquarello},
  \citenamefont {Paulatto}, \citenamefont {Sbraccia}, \citenamefont {Scandolo},
  \citenamefont {Sclauzero}, \citenamefont {Seitsonen}, \citenamefont
  {Smogunov}, \citenamefont {Umari},\ and\ \citenamefont
  {Wentzcovitch}}]{QE-2009}%
  \BibitemOpen
  \bibfield  {author} {\bibinfo {author} {\bibfnamefont {P.}~\bibnamefont
  {Giannozzi}}, \bibinfo {author} {\bibfnamefont {S.}~\bibnamefont {Baroni}},
  \bibinfo {author} {\bibfnamefont {N.}~\bibnamefont {Bonini}}, \bibinfo
  {author} {\bibfnamefont {M.}~\bibnamefont {Calandra}}, \bibinfo {author}
  {\bibfnamefont {R.}~\bibnamefont {Car}}, \bibinfo {author} {\bibfnamefont
  {C.}~\bibnamefont {Cavazzoni}}, \bibinfo {author} {\bibfnamefont
  {D.}~\bibnamefont {Ceresoli}}, \bibinfo {author} {\bibfnamefont {G.~L.}\
  \bibnamefont {Chiarotti}}, \bibinfo {author} {\bibfnamefont {M.}~\bibnamefont
  {Cococcioni}}, \bibinfo {author} {\bibfnamefont {I.}~\bibnamefont {Dabo}},
  \bibinfo {author} {\bibfnamefont {A.}~\bibnamefont {{Dal Corso}}}, \bibinfo
  {author} {\bibfnamefont {S.}~\bibnamefont {de~Gironcoli}}, \bibinfo {author}
  {\bibfnamefont {S.}~\bibnamefont {Fabris}}, \bibinfo {author} {\bibfnamefont
  {G.}~\bibnamefont {Fratesi}}, \bibinfo {author} {\bibfnamefont
  {R.}~\bibnamefont {Gebauer}}, \bibinfo {author} {\bibfnamefont
  {U.}~\bibnamefont {Gerstmann}}, \bibinfo {author} {\bibfnamefont
  {C.}~\bibnamefont {Gougoussis}}, \bibinfo {author} {\bibfnamefont
  {A.}~\bibnamefont {Kokalj}}, \bibinfo {author} {\bibfnamefont
  {M.}~\bibnamefont {Lazzeri}}, \bibinfo {author} {\bibfnamefont
  {L.}~\bibnamefont {Martin-Samos}}, \bibinfo {author} {\bibfnamefont
  {N.}~\bibnamefont {Marzari}}, \bibinfo {author} {\bibfnamefont
  {F.}~\bibnamefont {Mauri}}, \bibinfo {author} {\bibfnamefont
  {R.}~\bibnamefont {Mazzarello}}, \bibinfo {author} {\bibfnamefont
  {S.}~\bibnamefont {Paolini}}, \bibinfo {author} {\bibfnamefont
  {A.}~\bibnamefont {Pasquarello}}, \bibinfo {author} {\bibfnamefont
  {L.}~\bibnamefont {Paulatto}}, \bibinfo {author} {\bibfnamefont
  {C.}~\bibnamefont {Sbraccia}}, \bibinfo {author} {\bibfnamefont
  {S.}~\bibnamefont {Scandolo}}, \bibinfo {author} {\bibfnamefont
  {G.}~\bibnamefont {Sclauzero}}, \bibinfo {author} {\bibfnamefont {A.~P.}\
  \bibnamefont {Seitsonen}}, \bibinfo {author} {\bibfnamefont {A.}~\bibnamefont
  {Smogunov}}, \bibinfo {author} {\bibfnamefont {P.}~\bibnamefont {Umari}}, \
  and\ \bibinfo {author} {\bibfnamefont {R.~M.}\ \bibnamefont {Wentzcovitch}},\
  }\href {http://www.quantum-espresso.org} {\bibfield  {journal} {\bibinfo
  {journal} {Journal of Physics: Condensed Matter}\ }\textbf {\bibinfo {volume}
  {21}},\ \bibinfo {pages} {395502 (19pp)} (\bibinfo {year}
  {2009})}\BibitemShut {NoStop}%
\bibitem [{\citenamefont {Clark}\ \emph {et~al.}(2005)\citenamefont {Clark},
  \citenamefont {Segall}, \citenamefont {Pickard}, \citenamefont {Hasnip},
  \citenamefont {Probert}, \citenamefont {Refson},\ and\ \citenamefont
  {Payne}}]{CASTEP}%
  \BibitemOpen
  \bibfield  {author} {\bibinfo {author} {\bibfnamefont {S.~J.}\ \bibnamefont
  {Clark}}, \bibinfo {author} {\bibfnamefont {M.~D.}\ \bibnamefont {Segall}},
  \bibinfo {author} {\bibfnamefont {C.~J.}\ \bibnamefont {Pickard}}, \bibinfo
  {author} {\bibfnamefont {P.~J.}\ \bibnamefont {Hasnip}}, \bibinfo {author}
  {\bibfnamefont {M.~J.}\ \bibnamefont {Probert}}, \bibinfo {author}
  {\bibfnamefont {K.}~\bibnamefont {Refson}}, \ and\ \bibinfo {author}
  {\bibfnamefont {M.~C.}\ \bibnamefont {Payne}},\ }\href@noop {} {\bibfield
  {journal} {\bibinfo  {journal} {Zeitschrift f{\"u}r Kristallographie}\
  }\textbf {\bibinfo {volume} {220}},\ \bibinfo {pages} {567} (\bibinfo {year}
  {2005})}\BibitemShut {NoStop}%
\bibitem [{\citenamefont {Kresse}\ and\ \citenamefont
  {Hafner}(1993)}]{PhysRevB.47.558}%
  \BibitemOpen
  \bibfield  {author} {\bibinfo {author} {\bibfnamefont {G.}~\bibnamefont
  {Kresse}}\ and\ \bibinfo {author} {\bibfnamefont {J.}~\bibnamefont
  {Hafner}},\ }\href {\doibase 10.1103/PhysRevB.47.558} {\bibfield  {journal}
  {\bibinfo  {journal} {Phys. Rev. B}\ }\textbf {\bibinfo {volume} {47}},\
  \bibinfo {pages} {558} (\bibinfo {year} {1993})}\BibitemShut {NoStop}%
\bibitem [{\citenamefont {Kresse}\ and\ \citenamefont
  {Furthm\"uller}(1996)}]{PhysRevB.54.11169}%
  \BibitemOpen
  \bibfield  {author} {\bibinfo {author} {\bibfnamefont {G.}~\bibnamefont
  {Kresse}}\ and\ \bibinfo {author} {\bibfnamefont {J.}~\bibnamefont
  {Furthm\"uller}},\ }\href {\doibase 10.1103/PhysRevB.54.11169} {\bibfield
  {journal} {\bibinfo  {journal} {Phys. Rev. B}\ }\textbf {\bibinfo {volume}
  {54}},\ \bibinfo {pages} {11169} (\bibinfo {year} {1996})}\BibitemShut
  {NoStop}%
\bibitem [{\citenamefont {Gonze}(1997)}]{PhysRevB.55.10337}%
  \BibitemOpen
  \bibfield  {author} {\bibinfo {author} {\bibfnamefont {X.}~\bibnamefont
  {Gonze}},\ }\href {\doibase 10.1103/PhysRevB.55.10337} {\bibfield  {journal}
  {\bibinfo  {journal} {Phys. Rev. B}\ }\textbf {\bibinfo {volume} {55}},\
  \bibinfo {pages} {10337} (\bibinfo {year} {1997})}\BibitemShut {NoStop}%
\bibitem [{\citenamefont {Payne}\ \emph {et~al.}(1992)\citenamefont {Payne},
  \citenamefont {Teter}, \citenamefont {Allan}, \citenamefont {Arias},\ and\
  \citenamefont {Joannopoulos}}]{RevModPhys.64.1045}%
  \BibitemOpen
  \bibfield  {author} {\bibinfo {author} {\bibfnamefont {M.~C.}\ \bibnamefont
  {Payne}}, \bibinfo {author} {\bibfnamefont {M.~P.}\ \bibnamefont {Teter}},
  \bibinfo {author} {\bibfnamefont {D.~C.}\ \bibnamefont {Allan}}, \bibinfo
  {author} {\bibfnamefont {T.~A.}\ \bibnamefont {Arias}}, \ and\ \bibinfo
  {author} {\bibfnamefont {J.~D.}\ \bibnamefont {Joannopoulos}},\ }\href
  {\doibase 10.1103/RevModPhys.64.1045} {\bibfield  {journal} {\bibinfo
  {journal} {Rev. Mod. Phys.}\ }\textbf {\bibinfo {volume} {64}},\ \bibinfo
  {pages} {1045} (\bibinfo {year} {1992})}\BibitemShut {NoStop}%
\bibitem [{\citenamefont {Bl\"ochl}(1994)}]{PhysRevB.50.17953}%
  \BibitemOpen
  \bibfield  {author} {\bibinfo {author} {\bibfnamefont {P.~E.}\ \bibnamefont
  {Bl\"ochl}},\ }\href {\doibase 10.1103/PhysRevB.50.17953} {\bibfield
  {journal} {\bibinfo  {journal} {Phys. Rev. B}\ }\textbf {\bibinfo {volume}
  {50}},\ \bibinfo {pages} {17953} (\bibinfo {year} {1994})}\BibitemShut
  {NoStop}%
\bibitem [{\citenamefont {Kresse}\ and\ \citenamefont
  {Joubert}(1999)}]{PhysRevB.59.1758}%
  \BibitemOpen
  \bibfield  {author} {\bibinfo {author} {\bibfnamefont {G.}~\bibnamefont
  {Kresse}}\ and\ \bibinfo {author} {\bibfnamefont {D.}~\bibnamefont
  {Joubert}},\ }\href {\doibase 10.1103/PhysRevB.59.1758} {\bibfield  {journal}
  {\bibinfo  {journal} {Phys. Rev. B}\ }\textbf {\bibinfo {volume} {59}},\
  \bibinfo {pages} {1758} (\bibinfo {year} {1999})}\BibitemShut {NoStop}%
\bibitem [{\citenamefont {Perdew}\ and\ \citenamefont
  {Zunger}(1981)}]{PhysRevB.23.5048}%
  \BibitemOpen
  \bibfield  {author} {\bibinfo {author} {\bibfnamefont {J.~P.}\ \bibnamefont
  {Perdew}}\ and\ \bibinfo {author} {\bibfnamefont {A.}~\bibnamefont
  {Zunger}},\ }\href {\doibase 10.1103/PhysRevB.23.5048} {\bibfield  {journal}
  {\bibinfo  {journal} {Phys. Rev. B}\ }\textbf {\bibinfo {volume} {23}},\
  \bibinfo {pages} {5048} (\bibinfo {year} {1981})}\BibitemShut {NoStop}%
\bibitem [{\citenamefont {Ceperley}\ and\ \citenamefont
  {Alder}(1980)}]{PhysRevLett.45.566}%
  \BibitemOpen
  \bibfield  {author} {\bibinfo {author} {\bibfnamefont {D.~M.}\ \bibnamefont
  {Ceperley}}\ and\ \bibinfo {author} {\bibfnamefont {B.~J.}\ \bibnamefont
  {Alder}},\ }\href {\doibase 10.1103/PhysRevLett.45.566} {\bibfield  {journal}
  {\bibinfo  {journal} {Phys. Rev. Lett.}\ }\textbf {\bibinfo {volume} {45}},\
  \bibinfo {pages} {566} (\bibinfo {year} {1980})}\BibitemShut {NoStop}%
\bibitem [{\citenamefont {Vanderbilt}(2000)}]{VANDERBILT2000147}%
  \BibitemOpen
  \bibfield  {author} {\bibinfo {author} {\bibfnamefont {D.}~\bibnamefont
  {Vanderbilt}},\ }\href {\doibase
  https://doi.org/10.1016/S0022-3697(99)00273-5} {\bibfield  {journal}
  {\bibinfo  {journal} {Journal of Physics and Chemistry of Solids}\ }\textbf
  {\bibinfo {volume} {61}},\ \bibinfo {pages} {147 } (\bibinfo {year}
  {2000})}\BibitemShut {NoStop}%
\bibitem [{\citenamefont {Gajdo\ifmmode~\check{s}\else \v{s}\fi{}}\ \emph
  {et~al.}(2006)\citenamefont {Gajdo\ifmmode~\check{s}\else \v{s}\fi{}},
  \citenamefont {Hummer}, \citenamefont {Kresse}, \citenamefont
  {Furthm\"uller},\ and\ \citenamefont {Bechstedt}}]{PhysRevB.73.045112}%
  \BibitemOpen
  \bibfield  {author} {\bibinfo {author} {\bibfnamefont {M.}~\bibnamefont
  {Gajdo\ifmmode~\check{s}\else \v{s}\fi{}}}, \bibinfo {author} {\bibfnamefont
  {K.}~\bibnamefont {Hummer}}, \bibinfo {author} {\bibfnamefont
  {G.}~\bibnamefont {Kresse}}, \bibinfo {author} {\bibfnamefont
  {J.}~\bibnamefont {Furthm\"uller}}, \ and\ \bibinfo {author} {\bibfnamefont
  {F.}~\bibnamefont {Bechstedt}},\ }\href {\doibase 10.1103/PhysRevB.73.045112}
  {\bibfield  {journal} {\bibinfo  {journal} {Phys. Rev. B}\ }\textbf {\bibinfo
  {volume} {73}},\ \bibinfo {pages} {045112} (\bibinfo {year}
  {2006})}\BibitemShut {NoStop}%
\bibitem [{\citenamefont {Baroni}\ \emph {et~al.}(2001)\citenamefont {Baroni},
  \citenamefont {de~Gironcoli}, \citenamefont {Dal~Corso},\ and\ \citenamefont
  {Giannozzi}}]{RevModPhys.73.515}%
  \BibitemOpen
  \bibfield  {author} {\bibinfo {author} {\bibfnamefont {S.}~\bibnamefont
  {Baroni}}, \bibinfo {author} {\bibfnamefont {S.}~\bibnamefont
  {de~Gironcoli}}, \bibinfo {author} {\bibfnamefont {A.}~\bibnamefont
  {Dal~Corso}}, \ and\ \bibinfo {author} {\bibfnamefont {P.}~\bibnamefont
  {Giannozzi}},\ }\href {\doibase 10.1103/RevModPhys.73.515} {\bibfield
  {journal} {\bibinfo  {journal} {Rev. Mod. Phys.}\ }\textbf {\bibinfo {volume}
  {73}},\ \bibinfo {pages} {515} (\bibinfo {year} {2001})}\BibitemShut
  {NoStop}%
\bibitem [{\citenamefont {Wu}\ \emph {et~al.}(2005)\citenamefont {Wu},
  \citenamefont {Vanderbilt},\ and\ \citenamefont
  {Hamann}}]{PhysRevB.72.035105}%
  \BibitemOpen
  \bibfield  {author} {\bibinfo {author} {\bibfnamefont {X.}~\bibnamefont
  {Wu}}, \bibinfo {author} {\bibfnamefont {D.}~\bibnamefont {Vanderbilt}}, \
  and\ \bibinfo {author} {\bibfnamefont {D.~R.}\ \bibnamefont {Hamann}},\
  }\href {\doibase 10.1103/PhysRevB.72.035105} {\bibfield  {journal} {\bibinfo
  {journal} {Phys. Rev. B}\ }\textbf {\bibinfo {volume} {72}},\ \bibinfo
  {pages} {035105} (\bibinfo {year} {2005})}\BibitemShut {NoStop}%
\bibitem [{\citenamefont {Perdew}\ \emph {et~al.}(1996)\citenamefont {Perdew},
  \citenamefont {Burke},\ and\ \citenamefont
  {Ernzerhof}}]{PhysRevLett.77.3865}%
  \BibitemOpen
  \bibfield  {author} {\bibinfo {author} {\bibfnamefont {J.~P.}\ \bibnamefont
  {Perdew}}, \bibinfo {author} {\bibfnamefont {K.}~\bibnamefont {Burke}}, \
  and\ \bibinfo {author} {\bibfnamefont {M.}~\bibnamefont {Ernzerhof}},\ }\href
  {\doibase 10.1103/PhysRevLett.77.3865} {\bibfield  {journal} {\bibinfo
  {journal} {Phys. Rev. Lett.}\ }\textbf {\bibinfo {volume} {77}},\ \bibinfo
  {pages} {3865} (\bibinfo {year} {1996})}\BibitemShut {NoStop}%
\bibitem [{\citenamefont {Sciau}\ \emph {et~al.}(2004)\citenamefont {Sciau},
  \citenamefont {Kania}, \citenamefont {Dkhil}, \citenamefont {Suard},\ and\
  \citenamefont {Ratuszna}}]{Sciau_2004}%
  \BibitemOpen
  \bibfield  {author} {\bibinfo {author} {\bibfnamefont {P.}~\bibnamefont
  {Sciau}}, \bibinfo {author} {\bibfnamefont {A.}~\bibnamefont {Kania}},
  \bibinfo {author} {\bibfnamefont {B.}~\bibnamefont {Dkhil}}, \bibinfo
  {author} {\bibfnamefont {E.}~\bibnamefont {Suard}}, \ and\ \bibinfo {author}
  {\bibfnamefont {A.}~\bibnamefont {Ratuszna}},\ }\href {\doibase
  10.1088/0953-8984/16/16/004} {\bibfield  {journal} {\bibinfo  {journal}
  {Journal of Physics: Condensed Matter}\ }\textbf {\bibinfo {volume} {16}},\
  \bibinfo {pages} {2795} (\bibinfo {year} {2004})}\BibitemShut {NoStop}%
\bibitem [{\citenamefont {Richman}(1968)}]{Richman_1968}%
  \BibitemOpen
  \bibfield  {author} {\bibinfo {author} {\bibfnamefont {D.}~\bibnamefont
  {Richman}},\ }\href {\doibase 10.1149/1.2411483} {\bibfield  {journal}
  {\bibinfo  {journal} {Journal of The Electrochemical Society}\ }\textbf
  {\bibinfo {volume} {115}},\ \bibinfo {pages} {945} (\bibinfo {year}
  {1968})}\BibitemShut {NoStop}%
\bibitem [{\citenamefont {Lucovsky}\ \emph {et~al.}(1971)\citenamefont
  {Lucovsky}, \citenamefont {Martin},\ and\ \citenamefont
  {Burstein}}]{PhysRevB.4.1367}%
  \BibitemOpen
  \bibfield  {author} {\bibinfo {author} {\bibfnamefont {G.}~\bibnamefont
  {Lucovsky}}, \bibinfo {author} {\bibfnamefont {R.~M.}\ \bibnamefont
  {Martin}}, \ and\ \bibinfo {author} {\bibfnamefont {E.}~\bibnamefont
  {Burstein}},\ }\href {\doibase 10.1103/PhysRevB.4.1367} {\bibfield  {journal}
  {\bibinfo  {journal} {Phys. Rev. B}\ }\textbf {\bibinfo {volume} {4}},\
  \bibinfo {pages} {1367} (\bibinfo {year} {1971})}\BibitemShut {NoStop}%
\bibitem [{\citenamefont {Adachi}(2005)}]{Adachi2005}%
  \BibitemOpen
  \bibfield  {author} {\bibinfo {author} {\bibfnamefont {S.}~\bibnamefont
  {Adachi}},\ }\href@noop {} {\emph {\bibinfo {title} {\textit{Properties of
  Group-IV, III-V, and II-VI Semiconductors}}}}\ (\bibinfo  {publisher} {Wiley,
  Chichester, UK},\ \bibinfo {year} {2005})\BibitemShut {NoStop}%
\bibitem [{\citenamefont {Monemar}(1970)}]{MONEMAR19701295}%
  \BibitemOpen
  \bibfield  {author} {\bibinfo {author} {\bibfnamefont {B.}~\bibnamefont
  {Monemar}},\ }\href {\doibase https://doi.org/10.1016/0038-1098(70)90623-X}
  {\bibfield  {journal} {\bibinfo  {journal} {Solid State Communications}\
  }\textbf {\bibinfo {volume} {8}},\ \bibinfo {pages} {1295 } (\bibinfo {year}
  {1970})}\BibitemShut {NoStop}%
\bibitem [{\citenamefont {Ettenberg}\ and\ \citenamefont
  {Paff}(1970)}]{doi:10.1063/1.1658389}%
  \BibitemOpen
  \bibfield  {author} {\bibinfo {author} {\bibfnamefont {M.}~\bibnamefont
  {Ettenberg}}\ and\ \bibinfo {author} {\bibfnamefont {R.~J.}\ \bibnamefont
  {Paff}},\ }\href {\doibase 10.1063/1.1658389} {\bibfield  {journal} {\bibinfo
   {journal} {Journal of Applied Physics}\ }\textbf {\bibinfo {volume} {41}},\
  \bibinfo {pages} {3926} (\bibinfo {year} {1970})}\BibitemShut {NoStop}%
\bibitem [{\citenamefont {Spencer}\ \emph {et~al.}(1994)\citenamefont
  {Spencer}, \citenamefont {Ho}, \citenamefont {Men\'endez}, \citenamefont
  {Droopad}, \citenamefont {Fathollahnejad},\ and\ \citenamefont
  {Maracas}}]{PhysRevB.50.14125}%
  \BibitemOpen
  \bibfield  {author} {\bibinfo {author} {\bibfnamefont {G.~S.}\ \bibnamefont
  {Spencer}}, \bibinfo {author} {\bibfnamefont {A.~C.}\ \bibnamefont {Ho}},
  \bibinfo {author} {\bibfnamefont {J.}~\bibnamefont {Men\'endez}}, \bibinfo
  {author} {\bibfnamefont {R.}~\bibnamefont {Droopad}}, \bibinfo {author}
  {\bibfnamefont {H.}~\bibnamefont {Fathollahnejad}}, \ and\ \bibinfo {author}
  {\bibfnamefont {G.~N.}\ \bibnamefont {Maracas}},\ }\href {\doibase
  10.1103/PhysRevB.50.14125} {\bibfield  {journal} {\bibinfo  {journal} {Phys.
  Rev. B}\ }\textbf {\bibinfo {volume} {50}},\ \bibinfo {pages} {14125}
  (\bibinfo {year} {1994})}\BibitemShut {NoStop}%
\bibitem [{\citenamefont {Guyaux}\ \emph {et~al.}(1992)\citenamefont {Guyaux},
  \citenamefont {Degiovanni}, \citenamefont {Thiry}, \citenamefont {Sporken},\
  and\ \citenamefont {Caudano}}]{GUYAUX1992697}%
  \BibitemOpen
  \bibfield  {author} {\bibinfo {author} {\bibfnamefont {J.}~\bibnamefont
  {Guyaux}}, \bibinfo {author} {\bibfnamefont {A.}~\bibnamefont {Degiovanni}},
  \bibinfo {author} {\bibfnamefont {P.}~\bibnamefont {Thiry}}, \bibinfo
  {author} {\bibfnamefont {R.}~\bibnamefont {Sporken}}, \ and\ \bibinfo
  {author} {\bibfnamefont {R.}~\bibnamefont {Caudano}},\ }\href {\doibase
  https://doi.org/10.1016/0169-4332(92)90324-Q} {\bibfield  {journal} {\bibinfo
   {journal} {Applied Surface Science}\ }\textbf {\bibinfo {volume} {56-58}},\
  \bibinfo {pages} {697 } (\bibinfo {year} {1992})}\BibitemShut {NoStop}%
\bibitem [{\citenamefont {Giesecke}\ and\ \citenamefont
  {Pfister}(1958)}]{Giesecke:a02256}%
  \BibitemOpen
  \bibfield  {author} {\bibinfo {author} {\bibfnamefont {G.}~\bibnamefont
  {Giesecke}}\ and\ \bibinfo {author} {\bibfnamefont {H.}~\bibnamefont
  {Pfister}},\ }\href {\doibase 10.1107/S0365110X58000979} {\bibfield
  {journal} {\bibinfo  {journal} {Acta Crystallographica}\ }\textbf {\bibinfo
  {volume} {11}},\ \bibinfo {pages} {369} (\bibinfo {year} {1958})}\BibitemShut
  {NoStop}%
\bibitem [{\citenamefont {Hass}\ and\ \citenamefont
  {Henvis}(1962)}]{HASS19621099}%
  \BibitemOpen
  \bibfield  {author} {\bibinfo {author} {\bibfnamefont {M.}~\bibnamefont
  {Hass}}\ and\ \bibinfo {author} {\bibfnamefont {B.}~\bibnamefont {Henvis}},\
  }\href {\doibase https://doi.org/10.1016/0022-3697(62)90127-0} {\bibfield
  {journal} {\bibinfo  {journal} {Journal of Physics and Chemistry of Solids}\
  }\textbf {\bibinfo {volume} {23}},\ \bibinfo {pages} {1099 } (\bibinfo {year}
  {1962})}\BibitemShut {NoStop}%
\bibitem [{\citenamefont {Charra}\ and\ \citenamefont
  {Gurzadyan}(2000)}]{LandoltBornstein2000:sm_lbs_978-3-540-37124-3_40}%
  \BibitemOpen
  \bibfield  {author} {\bibinfo {author} {\bibfnamefont {F.}~\bibnamefont
  {Charra}}\ and\ \bibinfo {author} {\bibfnamefont {G.}~\bibnamefont
  {Gurzadyan}},\ }\href {\doibase 10.1007/10134958_40} {\emph {\bibinfo {title}
  {\textit{Nonlinear Dielectric Susceptibilities: Datasheet from
  Landolt-B{\"o}rnstein-Group III Condensed Matter, Volume 30B}}}},\ edited by\
  \bibinfo {editor} {\bibfnamefont {D.~F.}\ \bibnamefont {Nelson}}\ (\bibinfo
  {publisher} {Springer-Verlag Berlin Heidelberg},\ \bibinfo {year}
  {2000})\BibitemShut {NoStop}%
\bibitem [{\citenamefont {Turner}\ and\ \citenamefont
  {Reese}(1962)}]{PhysRev.127.126}%
  \BibitemOpen
  \bibfield  {author} {\bibinfo {author} {\bibfnamefont {W.~J.}\ \bibnamefont
  {Turner}}\ and\ \bibinfo {author} {\bibfnamefont {W.~E.}\ \bibnamefont
  {Reese}},\ }\href {\doibase 10.1103/PhysRev.127.126} {\bibfield  {journal}
  {\bibinfo  {journal} {Phys. Rev.}\ }\textbf {\bibinfo {volume} {127}},\
  \bibinfo {pages} {126} (\bibinfo {year} {1962})}\BibitemShut {NoStop}%
\bibitem [{\citenamefont {Singh}(1986)}]{Singh1986_book}%
  \BibitemOpen
  \bibfield  {author} {\bibinfo {author} {\bibfnamefont {S.}~\bibnamefont
  {Singh}},\ }\href@noop {} {\emph {\bibinfo {title} {\textit{Handbook of Laser
  Science and Technology}}}}\ (\bibinfo  {publisher} {edited by M.J. Weber
  (CRC, Boca Raton, FL), Vol. III},\ \bibinfo {year} {1986})\BibitemShut
  {NoStop}%
\bibitem [{\citenamefont {Corso}\ \emph {et~al.}(1996)\citenamefont {Corso},
  \citenamefont {Mauri},\ and\ \citenamefont {Rubio}}]{PhysRevB.53.15638}%
  \BibitemOpen
  \bibfield  {author} {\bibinfo {author} {\bibfnamefont {A.~D.}\ \bibnamefont
  {Corso}}, \bibinfo {author} {\bibfnamefont {F.}~\bibnamefont {Mauri}}, \ and\
  \bibinfo {author} {\bibfnamefont {A.}~\bibnamefont {Rubio}},\ }\href
  {\doibase 10.1103/PhysRevB.53.15638} {\bibfield  {journal} {\bibinfo
  {journal} {Phys. Rev. B}\ }\textbf {\bibinfo {volume} {53}},\ \bibinfo
  {pages} {15638} (\bibinfo {year} {1996})}\BibitemShut {NoStop}%
\bibitem [{\citenamefont {Levin}\ \emph {et~al.}(2009)\citenamefont {Levin},
  \citenamefont {Krayzman}, \citenamefont {Woicik}, \citenamefont
  {Karapetrova}, \citenamefont {Proffen}, \citenamefont {Tucker},\ and\
  \citenamefont {Reaney}}]{PhysRevB.79.104113}%
  \BibitemOpen
  \bibfield  {author} {\bibinfo {author} {\bibfnamefont {I.}~\bibnamefont
  {Levin}}, \bibinfo {author} {\bibfnamefont {V.}~\bibnamefont {Krayzman}},
  \bibinfo {author} {\bibfnamefont {J.~C.}\ \bibnamefont {Woicik}}, \bibinfo
  {author} {\bibfnamefont {J.}~\bibnamefont {Karapetrova}}, \bibinfo {author}
  {\bibfnamefont {T.}~\bibnamefont {Proffen}}, \bibinfo {author} {\bibfnamefont
  {M.~G.}\ \bibnamefont {Tucker}}, \ and\ \bibinfo {author} {\bibfnamefont
  {I.~M.}\ \bibnamefont {Reaney}},\ }\href {\doibase
  10.1103/PhysRevB.79.104113} {\bibfield  {journal} {\bibinfo  {journal} {Phys.
  Rev. B}\ }\textbf {\bibinfo {volume} {79}},\ \bibinfo {pages} {104113}
  (\bibinfo {year} {2009})}\BibitemShut {NoStop}%
\bibitem [{\citenamefont {Yashima}\ \emph {et~al.}(2011)\citenamefont
  {Yashima}, \citenamefont {Matsuyama}, \citenamefont {Sano}, \citenamefont
  {Itoh}, \citenamefont {Tsuda},\ and\ \citenamefont {Fu}}]{Yashima_2011}%
  \BibitemOpen
  \bibfield  {author} {\bibinfo {author} {\bibfnamefont {M.}~\bibnamefont
  {Yashima}}, \bibinfo {author} {\bibfnamefont {S.}~\bibnamefont {Matsuyama}},
  \bibinfo {author} {\bibfnamefont {R.}~\bibnamefont {Sano}}, \bibinfo {author}
  {\bibfnamefont {M.}~\bibnamefont {Itoh}}, \bibinfo {author} {\bibfnamefont
  {K.}~\bibnamefont {Tsuda}}, \ and\ \bibinfo {author} {\bibfnamefont
  {D.}~\bibnamefont {Fu}},\ }\href {https://doi.org/10.1021/cm103389q}
  {\bibfield  {journal} {\bibinfo  {journal} {Chemistry of Materials}\ }\textbf
  {\bibinfo {volume} {23}},\ \bibinfo {pages} {1643} (\bibinfo {year}
  {2011})}\BibitemShut {NoStop}%
\bibitem [{\citenamefont {Chang}\ \emph {et~al.}(2012)\citenamefont {Chang},
  \citenamefont {Shang}, \citenamefont {Zhang}, \citenamefont {Yuan},\ and\
  \citenamefont {Feng}}]{doi:10.1111/j.1551-2916.2012.05392.x}%
  \BibitemOpen
  \bibfield  {author} {\bibinfo {author} {\bibfnamefont {H.}~\bibnamefont
  {Chang}}, \bibinfo {author} {\bibfnamefont {M.}~\bibnamefont {Shang}},
  \bibinfo {author} {\bibfnamefont {C.}~\bibnamefont {Zhang}}, \bibinfo
  {author} {\bibfnamefont {H.}~\bibnamefont {Yuan}}, \ and\ \bibinfo {author}
  {\bibfnamefont {S.}~\bibnamefont {Feng}},\ }\href {\doibase
  10.1111/j.1551-2916.2012.05392.x} {\bibfield  {journal} {\bibinfo  {journal}
  {Journal of the American Ceramic Society}\ }\textbf {\bibinfo {volume}
  {95}},\ \bibinfo {pages} {3673} (\bibinfo {year} {2012})}\BibitemShut
  {NoStop}%
\bibitem [{\citenamefont {Tian}\ \emph {et~al.}(2022)\citenamefont {Tian},
  \citenamefont {Song}, \citenamefont {Viola}, \citenamefont {Shi},
  \citenamefont {Li}, \citenamefont {Jin}, \citenamefont {Hu}, \citenamefont
  {Xu}, \citenamefont {Ge}, \citenamefont {Yan}, \citenamefont {Zhang},\ and\
  \citenamefont {Tarakina}}]{TianEtAl2022AgNbO3}%
  \BibitemOpen
  \bibfield  {author} {\bibinfo {author} {\bibfnamefont {Y.}~\bibnamefont
  {Tian}}, \bibinfo {author} {\bibfnamefont {P.}~\bibnamefont {Song}}, \bibinfo
  {author} {\bibfnamefont {G.}~\bibnamefont {Viola}}, \bibinfo {author}
  {\bibfnamefont {J.}~\bibnamefont {Shi}}, \bibinfo {author} {\bibfnamefont
  {J.}~\bibnamefont {Li}}, \bibinfo {author} {\bibfnamefont {L.}~\bibnamefont
  {Jin}}, \bibinfo {author} {\bibfnamefont {Q.}~\bibnamefont {Hu}}, \bibinfo
  {author} {\bibfnamefont {Y.}~\bibnamefont {Xu}}, \bibinfo {author}
  {\bibfnamefont {W.}~\bibnamefont {Ge}}, \bibinfo {author} {\bibfnamefont
  {Z.}~\bibnamefont {Yan}}, \bibinfo {author} {\bibfnamefont {D.}~\bibnamefont
  {Zhang}}, \ and\ \bibinfo {author} {\bibfnamefont {N.~V.}\ \bibnamefont
  {Tarakina}},\ }\href {\doibase 10.1039/D2TA00905F} {\bibfield  {journal}
  {\bibinfo  {journal} {Journal of Materials Chemistry A}\ }\textbf {\bibinfo
  {volume} {10}},\ \bibinfo {pages} {14747} (\bibinfo {year}
  {2022})}\BibitemShut {NoStop}%
\bibitem [{\citenamefont {Moriwake}\ \emph {et~al.}(2012)\citenamefont
  {Moriwake}, \citenamefont {Fisher}, \citenamefont {Kuwabara},\ and\
  \citenamefont {Fu}}]{Moriwake_2012}%
  \BibitemOpen
  \bibfield  {author} {\bibinfo {author} {\bibfnamefont {H.}~\bibnamefont
  {Moriwake}}, \bibinfo {author} {\bibfnamefont {C.~A.~J.}\ \bibnamefont
  {Fisher}}, \bibinfo {author} {\bibfnamefont {A.}~\bibnamefont {Kuwabara}}, \
  and\ \bibinfo {author} {\bibfnamefont {D.}~\bibnamefont {Fu}},\ }\href
  {\doibase 10.1143/jjap.51.09le02} {\bibfield  {journal} {\bibinfo  {journal}
  {Japanese Journal of Applied Physics}\ }\textbf {\bibinfo {volume} {51}},\
  \bibinfo {pages} {09LE02} (\bibinfo {year} {2012})}\BibitemShut {NoStop}%
\bibitem [{\citenamefont {Fu}\ \emph {et~al.}(2007)\citenamefont {Fu},
  \citenamefont {Endo}, \citenamefont {Taniguchi}, \citenamefont {Taniyama},\
  and\ \citenamefont {Itoh}}]{doi:10.1063/1.2751136}%
  \BibitemOpen
  \bibfield  {author} {\bibinfo {author} {\bibfnamefont {D.}~\bibnamefont
  {Fu}}, \bibinfo {author} {\bibfnamefont {M.}~\bibnamefont {Endo}}, \bibinfo
  {author} {\bibfnamefont {H.}~\bibnamefont {Taniguchi}}, \bibinfo {author}
  {\bibfnamefont {T.}~\bibnamefont {Taniyama}}, \ and\ \bibinfo {author}
  {\bibfnamefont {M.}~\bibnamefont {Itoh}},\ }\href {\doibase
  10.1063/1.2751136} {\bibfield  {journal} {\bibinfo  {journal} {Applied
  Physics Letters}\ }\textbf {\bibinfo {volume} {90}},\ \bibinfo {pages}
  {252907} (\bibinfo {year} {2007})}\BibitemShut {NoStop}%
\bibitem [{\citenamefont {Gao}\ \emph {et~al.}(2018)\citenamefont {Gao},
  \citenamefont {Zhao}, \citenamefont {Liu}, \citenamefont {Wang},
  \citenamefont {Zhang},\ and\ \citenamefont {Li}}]{doi:10.1111/jace.15780}%
  \BibitemOpen
  \bibfield  {author} {\bibinfo {author} {\bibfnamefont {J.}~\bibnamefont
  {Gao}}, \bibinfo {author} {\bibfnamefont {L.}~\bibnamefont {Zhao}}, \bibinfo
  {author} {\bibfnamefont {Q.}~\bibnamefont {Liu}}, \bibinfo {author}
  {\bibfnamefont {X.}~\bibnamefont {Wang}}, \bibinfo {author} {\bibfnamefont
  {S.}~\bibnamefont {Zhang}}, \ and\ \bibinfo {author} {\bibfnamefont {J.-F.}\
  \bibnamefont {Li}},\ }\href {\doibase 10.1111/jace.15780} {\bibfield
  {journal} {\bibinfo  {journal} {Journal of the American Ceramic Society}\
  }\textbf {\bibinfo {volume} {101}},\ \bibinfo {pages} {5443} (\bibinfo {year}
  {2018})}\BibitemShut {NoStop}%
\bibitem [{not()}]{note}%
  \BibitemOpen
  \href@noop {} {\enquote {\bibinfo {title} {\rm {P}rivate comuniation:
  optimized lattice parameters used in {Ref}.~\cite{doi:10.1063/1.4941319} are
  $a$=5.650 {$\AA$}, $b$=5.733 {$\AA$}, and $c$= 15.533 {$\AA$}.}}\
  }\BibitemShut {NoStop}%
\end{thebibliography}%

\end{document}